\documentclass[reprint, showpacs, superscriptaddress, aip, pof, longbibliography, twocolumn, 10pt]{revtex4-1}
\usepackage{graphicx}
\usepackage{color}
\usepackage{amssymb}
\usepackage{amsmath}
\usepackage{tabularx}
\usepackage{bm}
\usepackage{hyperref}
\usepackage{ltxtable}
\usepackage{natbib} 
\usepackage{longtable}
\usepackage{float}

\newcommand{\be}{\begin{equation}}
\newcommand{\ee}{\end{equation}}
\newcommand{\bse}{\begin{subequations}}
	\newcommand{\ese}{\end{subequations}}
\newcommand{\bea}{\begin{eqnarray}}
\newcommand{\eea}{\end{eqnarray}}
\newcommand{\ba}{\begin{array}}
	\newcommand{\ea}{\end{array}}

\usepackage[usenames,dvipsnames]{xcolor}
\hypersetup{
colorlinks,
citecolor=blue,
filecolor=blue,
linkcolor=blue,
urlcolor=blue}

\begin{document}

\title{\color{black}Revisiting Reynolds and Nusselt numbers in turbulent thermal convection}
\author{Shashwat Bhattacharya}
\email{shabhatt@iitk.ac.in}
\affiliation{Department of Mechanical Engineering, Indian Institute of Technology Kanpur, Kanpur 208016, India}
\author{Mahendra K. Verma}
\affiliation{Department of Physics, Indian Institute of Technology Kanpur, Kanpur 208016, India}
\author{{Ravi Samtaney}}
\affiliation{Mechanical Engineering, Division of Physical Science and Engineering, King Abdullah University of Science and Technology, Thuwal 23955, Saudi Arabia}

\date{\today}

\begin{abstract}
In this paper, we {\color{black}extend} Grossmann and Lohse's (GL) model [Phys. Rev. Lett. {\bf 86}, 3316 (2001)] for the predictions of Reynolds number (Re) and Nusselt number (Nu) in turbulent Rayleigh-B\'{e}nard convection (RBC). 
{\color{black}Towards this objective, we use functional forms for the prefactors of the dissipation rates in the bulk and the boundary layers. The functional forms arise due to inhibition of nonlinear interactions in the presence of walls and buoyancy compared to free turbulence, along with a deviation of viscous boundary layer profile from Prandtl-Blasius theory. } We perform 60 numerical runs on a three-dimensional unit box for a range  of Rayleigh numbers (Ra) and Prandtl numbers (Pr) and {\color{black} determine the aforementioned functional forms} using  machine learning. 
 The  {\color{black}revised predictions} are in better agreement with the past numerical and experimental results than those of the GL model, especially for extreme Prandtl numbers.

\end{abstract}
\maketitle

\section{Introduction} \label{sec:Introduction}
A classical problem in fluid dynamics is Rayleigh-B{\'e}nard convection (RBC), where a fluid {\color{black}is} enclosed between two horizontal walls with the bottom wall kept at a higher temperature than the top wall. RBC serves as a paradigm for many types of convective flows occurring in nature and in engineering applications. RBC is primarily governed by two parameters: the Rayleigh number Ra, which is the ratio of the buoyancy  and the dissipative force, and the Prandtl number Pr, which is the ratio of kinematic viscosity and thermal diffusivity of the fluid. In this paper, we derive {\color{black} a relation} to predict two important quantities -- the Nusselt number Nu and the Reynolds number Re, which are respective measures of large scale heat transport and velocity in turbulent RBC.

The dependence of Nu and Re on RBC's governing parameters (Ra and Pr) has been extensively studied in the literature.~\cite{Ahlers:RMP2009,Chilla:EPJE2012,Siggia:ARFM1994,Xia:TAML2013,Verma:book:BDF} \citet{Malkus:PRSA1954} proposed $\mathrm{Nu} \sim \mathrm{Ra}^{1/3}$ based on marginal stability theory. For very large Ra called ultimate regime, \citet{Kraichnan:PF1962Convection} deduced $\mathrm{Nu} \sim \sqrt{\mathrm{RaPr}}$, $\mathrm{Re} \sim \sqrt{\mathrm{Ra/Pr}}$ for $\mathrm{Pr} \leq 0.15$, and $\mathrm{Nu} \sim \sqrt{\mathrm{RaPr^{-1/2}}}$, $\mathrm{Re} \sim \sqrt{\mathrm{Ra/Pr^{3/2}}}$ for $0.15 < \mathrm{Pr} \leq 1$, with logarithmic corrections. Subsequently, \citet{Castaing:JFM1989}  argued that $\mathrm{Nu}\sim \mathrm{Ra}^{2/7}$ and $\mathrm{Re} \sim \mathrm{Ra}^{3/7}$ based on the existence of a mixing zone in the central region of the RBC cell where hot rising plumes meet mildly warm fluid. \citet{Castaing:JFM1989} also deduced that $\mathrm{Re}^\omega \sim \mathrm{Ra}^{1/2}$, where $\mathrm{Re}^\omega$ is Reynolds number based on the frequency $\omega$ of torsional azimuthal oscillations of the large scale wind in RBC.  Later, \citet{Shraiman:PRA1990} derived that $\mathrm{Nu} \sim \mathrm{Ra}^{2/7}\mathrm{Pr}^{-1/7}$ and $\mathrm{Re} \sim \mathrm{Ra}^{3/7}\mathrm{Pr}^{-5/7}$ (with logarithmic corrections) using the properties of boundary layers. They also derived exact relations between Nu and the viscous and thermal dissipation rates.

Many experiments and simulations of RBC have been performed to obtain the scaling of Nu and Re. These studies also revealed a power-law scaling of Nu and Re as $\mathrm{Nu} \sim \mathrm{Ra}^{\alpha}\mathrm{Pr}^{\beta}$ and $\mathrm{Re} \sim \mathrm{Ra}^{\gamma}\mathrm{Pr}^{\delta}$. For the scaling of Nu, the exponent $\alpha$ ranges from 1/4 for $\mathrm{Pr} \ll 1$ to approximately 1/3 for $\mathrm{Pr} \gtrsim 1$,~\cite{Cioni:JFM1997,Scheel:PRF2017,Rossby:JFM1969,Takeshita:PRL1996,Ashkenazi:PRL1999Nu,Castaing:JFM1989,Chavanne:PRL1997,Horn:JFM2013,Emran:JFM2008,Wagner:PF2013,Kaczorowski:JFM2013,Niemela:JFM2003,Funfschilling:JFM2005,Pandey:PF2016,Pandey:PRE2016,Pandey:PRE2014,Stevens:JFM2010,Xu:PF2019,Dong:PF2020,Madanan:PF2020,Vial:PF2017} and $\beta$ from approximately zero for $\mathrm{Pr}\gtrsim 1$ to $0.14$ for $\mathrm{Pr} \ll 1$~\cite{Xia:PRL2002,Verzicco:JFM1999}. Thus, Nu has a relatively weaker dependence on Pr. For the scaling of Re, the exponent $\gamma$ was observed to be approximately $2/5$ for $\mathrm{Pr}\ll 1$, $1/2$ for $\mathrm{Pr} \sim 1$, and $3/5$ for $\mathrm{Pr} \gg 1$;~\cite{Chavanne:PRL1997,Castaing:JFM1989,Emran:JFM2008,Niemela:JFM2001,Pandey:PF2016,Pandey:PRE2016,Wagner:PF2013,Lam:PRE2002,Verma:PRE2012,Scheel:PRF2017,Cioni:JFM1997,Pandey:PF2016,Verma:PRE2012,Horn:JFM2013,Lam:PRE2002,Pandey:PF2016,Pandey:PRE2016,Pandey:PRE2014,Silano:JFM2010} and $\delta$ has been observed to range from $ -0.7 $ for $\mathrm{Pr} \lesssim 1$ to $ -0.95 $ for $\mathrm{Pr} \gg 1$.~\cite{Xia:PRL2002,Brown:JSM2007} A careful examination of the results of the above references reveal that the above exponents also depend on the regime of Ra as well. The ultimate regime, characterized by $\mathrm{Nu} \sim \sqrt{\mathrm{Ra}}$, has been observed in simulations of RBC with periodic boundary conditions,~\cite{Verma:PRE2012,Lohse:PRL2003} in free convection with density gradient,~\cite{He:NJP2012,Pawar:PF2016,Pawar:PRF2016} and in convection with only lateral walls.~\cite{Schmidt:JFM2012} Using numerical simulations, \citet{Calzavarini:PF2005} showed that $\mathrm{Re} \sim \mathrm{Pr}^{1/2}$ and $\mathrm{Nu} \sim \mathrm{Pr}^{1/2}$ for convection with periodic walls. However, some doubts have been raised on the ultimate scaling observed in RBC with periodic walls because of the presence of elevator modes in the system.~\cite{Calzavarini:PRE2006,Doering:JFM2019} Some experiments and simulations of RBC with non-periodic walls and very large Ra ($\sim 10^{15}$) have reported a possible transition to the ultimate regime;~\cite{Chavanne:PRL1997,Roche:PRE2001,Ahlers:NJP2009,He:NJP2012} however, some others~\cite{Niemela:Nature2000,Iyer:PNAS2020} argue against such transition. 

The above studies show that the scaling of Re and Nu depends on the regime of Ra and Pr, highlighting the need for a unified model that encompasses all the regimes. Grossmann and Lohse~\cite{Grossmann:JFM2000,Grossmann:PRL2001,Grossmann:PRE2002,Grossmann:JFM2003} constructed one such model, henceforth referred to as GL model. To derive this model, \citet{Grossmann:JFM2000,Grossmann:PRL2001} substituted the bulk and the boundary layer contributions of viscous and thermal dissipation rates in the exact relations of \citet{Shraiman:PRA1990}. The bulk and the boundary layer contributions were written in terms of Re, Nu, Ra, and Pr using the properties of boundary layers (Prandtl-Blasius theory)~\cite{Landau:book:Fluid} and those of hydrodynamic and passive scalar turbulence in the bulk. Finally, using additional crossover functions, \citet{Grossmann:PRL2001} obtained a system of equations for Re and Nu in terms of Ra, Pr, and four coefficients that were determined using inputs from experimental data.~\cite{Stevens:JFM2013}  Using the momentum equation of RBC,  Pandey  et al.~\cite{Pandey:PF2016,Pandey:PRE2016} constructed a model to predict the Reynolds number as a function of {\color{black}Ra} and Pr. The predictions of \citet{Kraichnan:PF1962Convection}, \citet{Castaing:JFM1989}, and \citet{Shraiman:PRA1990}  are limiting cases of the GL model.

The GL model has been quite successful in predicting large scale velocity and heat transport in many experiments and simulations. However, it does not capture large Pr convection very accurately~\cite{Verma:book:BDF} and has been reported to under-predict the Reynolds number~\cite{Ahlers:RMP2009}. Note that the scaling exponent for Re has a longer range (0.40 to 0.60) compared to that for Nu (0.25 to 0.33); hence the predictions for Re are more sensitive to modeling parameters. Further, the GL model is based on certain assumptions that are not valid for RBC. For example, the model assumes that  the viscous and the thermal dissipation rate in the bulk scale as $U^3/d$ and $U\Delta^2/d$ (for $\mathrm{Pr} \lesssim 1$) respectively, as in passive scalar turbulence with open boundaries.~\cite{Lesieur:book:Turbulence,Verma:book:ET} Here $U$ is the large-scale velocity, and $\Delta$ and $d$ are respectively the temperature difference and the distance between the top and bottom walls. However, subsequent studies of RBC have shown that the aforementioned viscous and the thermal dissipation rates in the bulk are suppressed by approximately $\mathrm{Ra}^{-0.2}$ for $\mathrm{Pr} \sim 1$.~\cite{Verzicco:JFM2003,Emran:JFM2008,Pandey:PF2016,Pandey:PRE2016,Scheel:PRF2017,Bhattacharya:PF2018,Bhattacharya:PF2019} {\color{black}The above suppression is due to the inhibition of nonlinear interactions because of walls~\cite{Pandey:PF2016,Pandey:PRE2016} and buoyancy.~\cite{Bhattacharya:PF2019b}} Moreover, recent studies have revealed that the viscous boundary layer thickness in RBC considerably deviate from $\mathrm{Re}^{-1/2}$ as assumed in GL model.~\cite{Scheel:JFM2012,Shi:JFM2012,Bhattacharya:PF2018} 


In the present work, we {\color{black}address the above limitations of the GL model and propose a new relation for the Reynolds and Nusselt numbers involving a cubic polynomial equation for Re and Nu.}
For implementation of  the viscous and thermal dissipation rates in the bulk and the boundary layers, we employ machine learning  tools on 60 data sets that were obtained using numerical simulations of RBC. 
The {\color{black}new relation} rectifies some of the limitations of GL model, especially  for small and large Prandtl numbers.

The outline of the paper is as follows. In Sec~\ref{sec:Governing_Equations}, we discuss the governing equations of RBC and briefly explain the GL model. Then, we {\color{black}extend the GL framework by using functional forms for the prefactors of the dissipation rates in the bulk and boundary layers and incorporate the deviation in the scaling of viscous boundary layer thickness described earlier..} Simulation details are provided in Sec.~\ref{sec:Numerical_Method}. In Sec.~\ref{sec:Results}, we {\color{black}report} the scaling of boundary layer thicknesses and dissipation rates using our data, following which we describe the machine learning tools used  to determine  {\color{black}the aforementioned functional forms.} We also test the {\color{black}revised predictions} with experiments and numerical simulations, and {\color{black} compare them with those of the GL model.} We conclude in Sec.~\ref{sec:Conclusions}.   


\section{RBC equations and  the GL model} \label{sec:Governing_Equations}
We consider RBC under the Boussinesq approximation, whose governing equations are as follows~\cite{Chandrasekhar:book:Instability,Verma:book:BDF}:
\bea
\frac{\partial \mathbf{u}}{\partial t} + (\mathbf{u} \cdot \nabla) \mathbf{u} & = & -\nabla p/\rho_0 + \alpha g T \hat{z} + \nu \nabla^2 \mathbf{u}, \label{eq:momentum}\\
\frac{\partial T}{\partial t} + (\mathbf{u} \cdot \nabla) T & = & \kappa \nabla^2 T, \label{eq:energy}  \\
\nabla \cdot \mathbf{u} &=& 0, \label{eq:continuity} 
\eea  
where $\mathbf{u}$ and $p$ are the velocity and pressure fields respectively, $T$ is the temperature field, $\nu$ is the kinematic viscosity, $\kappa$ is the thermal diffusivity, $\alpha$ is the thermal expansion coefficient, $\rho_0$ is the mean density of the fluid, and $g$ is the acceleration due to gravity. 

Using $d$ as the length scale, $\sqrt{\alpha g \Delta d}$ as the velocity scale, and $\Delta$ as the temperature scale, we non-dimensionalize Eqs.~(\ref{eq:momentum})-(\ref{eq:continuity}) that yields
\bea
\frac{\partial \mathbf{u}}{\partial t} + \mathbf{u} \cdot \nabla \mathbf{u} &=& -\nabla p + T \hat{z} +  \mathrm{\sqrt{\frac{Pr}{Ra}}}\nabla^2 \mathbf{u}, \label{eq:NDMomentum} \\
\frac{\partial T}{\partial t} + \mathbf{u}\cdot \nabla T &=& \frac{1}{\sqrt{\mathrm{Ra Pr}}}\nabla^2 T, \label{eq:NDTheta}\\
\nabla \cdot \mathbf{u} &=& 0, \label{eq:NDContinuity}
\eea 
where $\mathrm{Ra} = \alpha g \Delta d^3/(\nu \kappa)$ is the Rayleigh number and $\mathrm{Pr} = \nu/\kappa$ is the Prandtl number.
The large scale velocity and heat transfer are quantified by two important non-dimensional quantities, namely, the Reynolds number (Re) and the Nusselt number (Nu). The Nusselt number Nu is the ratio of the total heat flux to the conductive heat flux, and is defined as $\mathrm{Nu} = 1 + \langle u_zT \rangle /(\kappa \Delta/d)$. The Reynolds number Re is defined as $\mathrm{Re}= Ud/\nu$,
where $U$ is the large-scale velocity. In our work, we will consider $U$ to be the root mean square (RMS) velocity, that is, $U = \sqrt{\langle u_x^2 + u_y^2 + u_z^2 \rangle}$, where $\langle \cdot \rangle$ represents the volume average.

The dissipation rate of kinetic and thermal energy, represented as $\epsilon_u$ and $\epsilon_T$ respectively, are important quantities in our study. These are defined as $\epsilon_u = 2 \nu \langle S_{ij}S_{ij} \rangle$, $\epsilon_T = \kappa \langle |\nabla T|^2 \rangle$, where $S_{ij}$ is the strain rate tensor.
\citet{Shraiman:PRA1990} derived two exact relations between Nu and the dissipation rates; these are
\begin{eqnarray}
\epsilon_u &=& \frac{\nu^3}{d^4}\mathrm{(Nu}-1)\frac{\mathrm{Ra}}{\mathrm{Pr}^2},
\label{eq:SS_Viscous} \\
\epsilon_T &=& \frac{\kappa \Delta^2}{d^2}\mathrm{Nu}.
\label{eq:SS_Thermal} 
\end{eqnarray}
The above relations will be the backbone of our present work.

Now, we will briefly summarize the GL model to predict Nu and Re. \citet{Grossmann:JFM2000,Grossmann:PRL2001} split the total viscous and thermal dissipation rates ($\tilde{D}_u = \epsilon_u V$ and $\tilde{D}_T = \epsilon_T V$ respectively, $V$ being the domain volume) into their bulk and boundary-layer contributions. 
 Thus,
\begin{eqnarray}
\tilde{D}_u &=& \tilde{D}_{u, \mathrm{bulk}} + \tilde{D}_{u, \mathrm{BL}},
\label{eq:TotalViscousDissipation} \\
\tilde{D}_T &=& \tilde{D}_{T, \mathrm{bulk}} + \tilde{D}_{T, \mathrm{BL}}.
\label{eq:TotalThermalDissipation}
\end{eqnarray}
The GL model assumes Prandtl-Blasius relation of $\delta_u \sim \mathrm{Re^{-1/2}}$ above a critical Reynolds number $\mathrm{Re}_c$ for viscous boundary layers, and $\delta_T = d/2\mathrm{Nu}$ for thermal boundary layers.   Here, $\delta_u$ and $\delta_T$ are the viscous and thermal boundary layer thicknesses respectively. For $\mathrm{Re} < \mathrm{Re}_c$, the viscous boundary layer is assumed to occupy the entire RBC cell. Using the above relations and the properties of hydrodynamic and passive scalar turbulence in the bulk (see Sec.~\ref{subsec:EuEt}),  \citet{Grossmann:JFM2000,Grossmann:PRL2001} deduced that
\begin{eqnarray}
\frac{1}{V}\tilde{D}_{u,\mathrm{bulk}} &\sim&  \frac{U^3}{d} =  c_1\frac{\nu^3}{d^4} \mathrm{Re^3},
\label{eq:U^3/d} \\
\frac{1}{V}\tilde{D}_{u,\mathrm{BL}} &\sim& \frac{\nu U^2}{\delta_u^2}\frac{\delta_u}{d} = c_2\frac{\nu^3}{d^4} \mathrm{Re^{2.5}},
\label{eq:nuU^2/d^2} \\
\frac{1}{V}\tilde{D}_{T,\mathrm{bulk}} &\sim&  \frac{U \Delta^2}{d} = c_3\frac{\kappa \Delta^2}{d^2} \mathrm{RePr},
\label{eq:UDelta^2/d} \\
\frac{1}{V}\tilde{D}_{T,\mathrm{BL}} &\sim&  \frac{\kappa \Delta^2}{\delta_T^2}\frac{\delta_T}{d} = c_4 \frac{\kappa \Delta^2}{d^2}\mathrm{Nu},
\label{eq:kappa Delta^2/d^2}
\end{eqnarray}
where $c_1$, $c_2$, $c_3$, and $c_4$ are constants. Note that for $\delta_u > \delta_T$ ($\mathrm{Pr} \gg 1$), \citet{Grossmann:JFM2000} modified Eq.~(\ref{eq:UDelta^2/d}) as
\begin{equation}
\frac{1}{V}\tilde{D}_{T,\mathrm{bulk}} \sim \frac{\delta_T}{\delta_u} \frac{U \Delta^2}{d} = c_3 \frac{\kappa \Delta^2}{d} \mathrm{Re}^{3/2}\mathrm{Pr}\mathrm{Nu}^{-1}.
\label{eq:UDelta^2/d_largePr}
\end{equation}

By approximating the dominant terms of Eq.~(\ref{eq:energy}) in the thermal boundary layers, \citet{Grossmann:JFM2000,Grossmann:PRL2001} further deduced that $\mathrm{Nu} \sim \mathrm{Re^{1/2}Pr^{1/2}}$ for $\delta_u<\delta_T$ and $\mathrm{Nu} \sim \mathrm{Re^{1/2}Pr^{1/3}}$ for $\delta_u>\delta_T$. To ensure smooth transition through different regimes of boundary layer thicknesses and Reynolds number, \citet{Grossmann:PRL2001} introduced two crossover functions, $f(x) = (1+x^4)^{-1/4}$ and $g(x) = x(1+x^4)^{-1/4}$, and applied them in the RHS of Eqs.~(\ref{eq:nuU^2/d^2})-(\ref{eq:UDelta^2/d_largePr}). Finally, \citet{Grossmann:PRL2001} put the modelling and splitting assumptions [Eqs.~(\ref{eq:TotalViscousDissipation})-(\ref{eq:UDelta^2/d_largePr})] together with the exact relations given by Eqs.~(\ref{eq:SS_Viscous}) and (\ref{eq:SS_Thermal}) to obtain the following set of equations for Nu and Re:
\begin{eqnarray}
\mathrm{(Nu-1)\frac{Ra}{Pr^2}} &=& c_1\mathrm{Re^3} + c_2 \frac{\mathrm{Re}^2}{g(\sqrt{\mathrm{Re}_c/\mathrm{Re}})}, \label{eq:GL1} \\
\mathrm{Nu} &=& c_3\mathrm{Pr Re} f \left[ \frac{2a\mathrm{Nu}}{\sqrt{\mathrm{Re}_c}} g\left(\sqrt{\frac{\mathrm{Re}_c}{\mathrm{Re}}} \right)\right] \nonumber \\
&& + c_4\sqrt{\mathrm{RePr}} \left \{
 f \left[ \frac{2a\mathrm{Nu}}{\sqrt{\mathrm{Re}_c}} g\left(\sqrt{\frac{\mathrm{Re}_c}{\mathrm{Re}}} \right)\right]
 \right \}^{1/2}. \nonumber \\ \label{eq:GL2}
\end{eqnarray}  
The values of the constants, obtained from experiments, are $c_1=1.38$, $c_2=8.05$, $c_3=0.0252$, $c_4=0.487$, $a=0.922$ and $\mathrm{Re}_c=3.401$ ~\cite{Stevens:JFM2013}. The above equations can be solved iteratively to obtain Re and Nu for given Ra and Pr.

Although the GL model has been quite successful in predicting Re and Nu, it  has certain deficiencies due to some assumptions that are invalid for RBC. First, recent studies reveal that the relation $\delta_u \sim \mathrm{Re}^{-1/2}$ for the viscous boundary layers is not strictly valid for RBC~\cite{Scheel:JFM2012,Shi:JFM2012,Bhattacharya:PF2018}. The viscous boundary layer thickness becomes a progressively weaker function of Re as Pr is increased~\cite{Breuer:PRE2004}. Thus, the relation given by Eq.~(\ref{eq:nuU^2/d^2}) is not  accurate. Second, as discussed earlier, studies have shown that  for $\mathrm{Pr}\sim 1$, the thermal and viscous dissipation rates in the bulk are suppressed relative to free turbulence:~\cite{Verzicco:JFM2003,Emran:JFM2008,Bhattacharya:PF2018,Bhattacharya:PF2019}
\begin{equation}
\frac{1}{V}\tilde{D}_{u,\mathrm{bulk}} \sim  \frac{U^3}{d}\mathrm{Ra}^{-0.18}, \quad 
\frac{1}{V}\tilde{D}_{T,\mathrm{bulk}} \sim \frac{U\Delta^2}{d}\mathrm{Ra}^{-0.20}. \nonumber
\end{equation}
Contrast the above relations with Eqs.~(\ref{eq:U^3/d}) and (\ref{eq:UDelta^2/d})~\cite{Bhattacharya:PF2018,Bhattacharya:PF2019} used in the GL model. This clearly signifies that $c_1$ and $c_3$ from Eqs.~(\ref{eq:U^3/d}) and (\ref{eq:UDelta^2/d}) cannot be treated as constants. Thus, it becomes imperative to study how $c_i$ varies with Ra and Pr in different regimes of RBC. 

We {\color{black}propose a modified relation for Re and Nu} by incorporating the aforementioned suppression of the total dissipation rates, as well as the modified law for the viscous boundary layers. Towards this objective, we make the following modifications to Eqs.~(\ref{eq:U^3/d})-(\ref{eq:kappa Delta^2/d^2}):
\begin{eqnarray}
\frac{1}{V}\tilde{D}_{u,\mathrm{bulk}} &=& f_1(\mathrm{Ra,Pr}) \frac{U^3}{d} =  f_1(\mathrm{Ra,Pr}) \frac{\nu^3}{d^4} \mathrm{Re^3},
\label{eq:fU^3/d} \\
\frac{1}{V}\tilde{D}_{u,\mathrm{BL}} &=& f_2(\mathrm{Ra,Pr}) \frac{\nu U^2}{\delta_u^2}\frac{\delta_u}{d} = f_2(\mathrm{Ra,Pr}) \frac{\nu^3}{d^4}\frac{d}{\delta_u} \mathrm{Re^{2}},
\label{eq:fnuU^2/d^2} \nonumber \\
\\
\frac{1}{V}\tilde{D}_{T,\mathrm{bulk}} &=&  f_3(\mathrm{Ra,Pr})\frac{U \Delta^2}{d} = f_3(\mathrm{Ra,Pr})\frac{\kappa \Delta^2}{d^2} \mathrm{RePr},
\label{eq:fUDelta^2/d} \nonumber \\
\\
\frac{1}{V}\tilde{D}_{T,\mathrm{BL}} &=&  f_4(\mathrm{Ra,Pr})\frac{\kappa \Delta^2}{\delta_T^2}\frac{\delta_T}{d} =f_4(\mathrm{Ra,Pr}) \frac{\kappa \Delta^2}{d^2}\mathrm{Nu}. \nonumber \\
\label{eq:fkappa Delta^2/d^2}
\end{eqnarray}
Note that we replaced the coefficients $c_i$ with functions $f_i(\mathrm{Ra,Pr})$. Further, we do not express $d/\delta_u$ in terms of Re in Eq.~(\ref{eq:fnuU^2/d^2}).
The above modified formulas are inserted in  the exact relations of \citet{Shraiman:PRA1990} that  leads to 
\begin{eqnarray}
(\mathrm{Nu}-1)\frac{\mathrm{Ra}}{\mathrm{Pr}^2} &=& f_1\mathrm{(Ra,Pr)}\mathrm{Re}^3 + f_2\mathrm{(Ra,Pr)} \frac{d}{\delta_u}\mathrm{Re}^2, \nonumber \\ \label{eq:Equation1} \\
\mathrm{Nu} &=& f_3 \mathrm{(Ra,Pr)}\mathrm{RePr} + 2f_4 \mathrm{(Ra,Pr)Nu}. \nonumber \\  \label{eq:Equation2}
\end{eqnarray}  
The functions $f_i(\mathrm{Ra,Pr})$ will be later determined using our simulation results. For the sake of brevity, we will skip the arguments within the parenthesis of $ f_i $'s  henceforth. 



Equations~(\ref{eq:Equation1}) and (\ref{eq:Equation2}) constitute a system of two equations with two unknowns (Re and Nu). To solve these equations, we will now reduce them to a cubic polynomial equation for Re by eliminating Nu. We rearrange Eq.~(\ref{eq:Equation2}) to obtain
\begin{equation}
\mathrm{Nu}=\frac{f_3}{1-2f_4}\mathrm{RePr}.
\label{eq:Nu_RePr}
\end{equation}
Substitution of Eq.~(\ref{eq:Nu_RePr}) in Eq.~(\ref{eq:Equation1}) yields the following cubic equation for Re:
\begin{equation}
f_1 \mathrm{Re}^3 + f_2 \frac{d}{\delta_u}\mathrm{Re}^2 - \frac{f_3}{1-2f_4}\mathrm{\frac{Ra}{Pr}Re} + \frac{\mathrm{Ra}}{\mathrm{Pr}^2}=0.
\label{eq:Cubic_Re}
\end{equation}
The above equation for Re can be solved for a given Ra and Pr once  $f_i$ and $\delta_u$ have been determined.  We  determine Nu using Eq.~(\ref{eq:Nu_RePr}) once Re has been computed.

Now, we will show that in the limit of viscous dissipation rate dominating in the bulk or in the boundary layers ($\tilde{D}_{u,\mathrm{bulk}} \gg \tilde{D}_{u,\mathrm{BL}}$ or vice-versa), {\color{black}Eqs.~(\ref{eq:Equation1},\ref{eq:Equation2}) are reduced to power-laws expressions for Re and Nu}. In the following discussion, we consider scaling for these limiting cases.

\subsection*{Case 1: $\tilde{D}_{u,\mathrm{bulk}} \gg \tilde{D}_{u,\mathrm{BL}}$}First, let us consider the case where viscous dissipation rate  in the bulk is dominant. This regime is expected for large Ra ($\gg 10^8$) or for small Pr ($ \ll 1$), where the boundary layers are thin. In this regime, $f_2 (d/\delta_u) \mathrm{Re^2} \ll f_1 \mathrm{Re^3}$. Assuming $\mathrm{Nu} \gg 1$, Eq.~(\ref{eq:Equation1}) reduces to
\begin{equation}
\mathrm{Nu \frac{Ra}{Pr^2}} \approx f_1 \mathrm{Re}^3.
\label{eq:Equation1_Turbulent}
\end{equation}
Using  Eqs.~(\ref{eq:Nu_RePr}, \ref{eq:Equation1_Turbulent}) we arrive at 
\begin{eqnarray}
\mathrm{Re} &=& \sqrt{\frac{f_3}{f_1(1-2f_4)} \mathrm{\frac{Ra}{Pr}}},
\label{eq:Re_Turbulent} \\
\mathrm{Nu} &=& \sqrt{\frac{1}{f_1}\left( \frac{f_3}{1-2f_4}\right)^3 \mathrm{RaPr}}.
\label{eq:Nu_Turbulent}
\end{eqnarray}
Note that $f_1$ and $f_3$ are expected to be constants and $f_4 \approx 0$ when the boundary layers are absent (as in a periodic box) or weak (as in the ultimate regime proposed by \citet{Kraichnan:PF1962Convection}). For this case, $\mathrm{Re} \sim \sqrt{\mathrm{Ra/Pr}}$ and $\mathrm{Nu} \sim \sqrt{\mathrm{RaPr}}$, consistent with the arguments of \citet{Kraichnan:PF1962Convection} for large Ra and small Pr. However, for RBC with walls, the relations for Re and Nu will deviate from the above relations because $f_1$ and $f_3$ are functions of Ra and Pr.

\subsection*{Case 2: $\tilde{D}_{u,\mathrm{BL}} \gg \tilde{D}_{u,\mathrm{bulk}}$}
Now, we consider the other extreme when the viscous dissipation rates  in the boundary layers are dominant, which is expected  for small Ra ($\ll 10^5$) or for large Pr ($\gg 7$).~\cite{Grossmann:JFM2000,Grossmann:PRL2001,Verzicco:JFM2003} In this regime, again assuming $\mathrm{Nu} \gg 1$, Eq.~(\ref{eq:Equation1}) reduces to
\begin{equation}
\mathrm{Nu \frac{Ra}{Pr^2}} \approx f_2 \frac{d}{\delta_u} \mathrm{Re}^2.
\label{eq:Equation1_Viscous}
\end{equation}
Using Eqs.~(\ref{eq:Nu_RePr}, \ref{eq:Equation1_Viscous}) we obtain
\begin{eqnarray}
\mathrm{Re} &=& \left \{ \frac{f_3}{f_2(1-2f_4)} \frac{\delta_u}{d} \right \}\mathrm{\frac{Ra}{Pr}},
\label{eq:Re_Viscous}\\
\mathrm{Nu} &=& \frac{1}{f_2}\frac{\delta_u}{d} \left( \frac{f_3}{1-2f_4} \right)^2  \mathrm{Ra}.
\label{eq:Nu_Viscous}
\end{eqnarray} 
We will examine these cases once we deduce the forms of $f_i$ using our numerical simulations.

We remark that the aspect ratio of the RBC cell also influences the scaling of Ra and Pr.~\cite{Grossmann:JFM2003} In the current work, we do not consider the effect of aspect ratio. We intend to include the aspect ratio dependence  in a future work.

In the next section, we will discuss  the simulation method.

\section{Simulation details} \label{sec:Numerical_Method}
\begin{table*}
	\caption{Details of our direct numerical simulations performed in a cubical box for $\mathrm{Pr} \leq 1$: the Prandtl number (Pr), the Rayleigh Number (Ra),  the grid size, the ratio of the Kolmogorov length scale~\cite{Batchelor:JFM1959_largeSc} ($\eta$) to the mesh width $\Delta x$, the number of grid points in viscous and thermal boundary layers ($N_{\mathrm{VBL}}$ and $N_{\mathrm{TBL}}$ respectively), the Reynolds number (Re), Nusselt number computed using $\langle u_zT \rangle$ and the exact relations given by Eqs.~(\ref{eq:SS_Viscous}) and (\ref{eq:SS_Thermal}) (Nu, $\mathrm{Nu}_u$, and $\mathrm{Nu}_T$ respectively), the ratio of the total viscous dissipation rate in the boundary layer ($\tilde{D}_{u,\mathrm{BL}}$) and that in the bulk ($\tilde{D}_{u,\mathrm{bulk}}$), the ratio of the total thermal dissipation rate in the boundary layer ($\tilde{D}_{T,\mathrm{BL}}$) and that in the bulk ($\tilde{D}_{T,\mathrm{bulk}}$), the number of non-dimensional time units ($t_\mathrm{ND}$) and snapshots over which the quantities are averaged.}
	\begin{ruledtabular}
		\begin{tabular}{c c c c c c c c c c c c c c}
			$\mathrm{Pr}$ & $\mathrm{Ra}$ & Grid size & $\eta/\Delta x$ & $N_{\mathrm{VBL}}$ & $N_{\mathrm{TBL}}$ & Re & Nu & $\mathrm{Nu}_u$ & $\mathrm{Nu}_T$ & $\frac{\tilde{D}_{u,\mathrm{BL}}}{\tilde{D}_{u,\mathrm{bulk}}}$ & $\frac{\tilde{D}_{T,\mathrm{BL}}}{\tilde{D}_{T,\mathrm{bulk}}}$ & $t_{\mathrm{ND}}$ & Snapshots\\
			\hline 
			$0.02$ & $\mathrm{5 \times 10^5}$ & $\mathrm{513^3}$ & 1.99 & $7$ & $58$ & $2440$ & $4.48$ & $4.54$ & $4.49$ & 0.751 & 2.90 & 95 &95\\
			$0.02$ & $\mathrm{1 \times 10^6}$ & $\mathrm{513^3}$ & 1.55 & $6$ & $46$ & $3200$ & 5.78 & 5.79 & 5.78 & 0.564 & 2.89 & 41 & 41\\
			$0.02$ & $\mathrm{2 \times 10^6}$ & $\mathrm{513^3}$ & 1.24 & $5$ & $38$ & $4290$ & 6.90 & 6.88 & 6.91 & 0.468 & 2.72 & 30 & 30\\
			$0.02$ & $\mathrm{5 \times 10^6}$ & $\mathrm{1025^3}$ & 1.81 & $7$ & $59$ & $6650$ & 8.85 & 9.18 & 8.89 & 0.381 & 2.68 & 7 & 71\\
			$0.02$ & $\mathrm{1 \times 10^7}$ & $\mathrm{1025^3}$ & 1.45 & $7$ & $48$ & $9420$ & 10.3 & 11.0 & 10.8 & 0.357 & 2.62 & 3 & 31\\ 
			$0.1$ & $\mathrm{5 \times 10^5}$ & $\mathrm{513^3}$ & 4.06 & $11$ & $43$ & $749$ & 6.11 & 6.11 & 6.11 & 0.911 & 2.89 & 107 & 107\\
			$0.1$ & $\mathrm{1 \times 10^6}$ & $\mathrm{513^3}$ & 3.23 & $9$ & $36$ & $1030$ & 7.34 & 7.39 & 7.35 & 0.787 & 2.71 & 66 & 66\\
			$0.1$ & $\mathrm{2 \times 10^6}$ & $\mathrm{513^3}$ & 2.58 & $7$ & $30$ & $1380$ & 8.85 & 8.83 & 8.86 & 0.646 & 2.66 & 88 & 88\\
			$0.1$ & $\mathrm{5 \times 10^6}$ & $\mathrm{513^3}$ & 1.91 & $6$ & $24$ & $2090$ & 11.3 & 11.4 & 11.3 & 0.539 & 2.63 & 83 & 83 \\
			$0.1$ & $\mathrm{1 \times 10^7}$ & $\mathrm{513^3}$ & 1.52 & $6$ & $20$ & $2870$ & 13.9 & 14.0 & 13.9 & 0.474 & 2.63 & 33 & 66\\
			$0.1$ & $\mathrm{2 \times 10^7}$ & $\mathrm{513^3}$ & 1.22 & $5$ & $17$ & $3870$ & 16.4 & 16.4 & 16.4 & 0.389 & 2.41 & 37 & 73\\
			$0.1$ & $\mathrm{5 \times 10^7}$ & $\mathrm{1025^3}$ & 1.83 & $7$ & $25$ & $6020$ & 20.8 & 20.8 & 21.3 & 0.337 & 2.22 & 12 & 12\\
			$0.1$ & $\mathrm{1 \times 10^8}$ & $\mathrm{1025^3}$ & 1.45 & $6$ & $21$ & $8140$ & 26.7 & 26.1 & 26.3 & 0.288 & 2.28 & 5 & 26\\ 
			$0.5$ & $\mathrm{1 \times 10^6}$ & $\mathrm{513^3}$ & 6.96 & $13$ & $32$ & $285$ & 8.38 & 8.36 & 8.37 & 1.01 & 3.25 & 71 & 71 \\
			$0.5$ & $\mathrm{3 \times 10^6}$ & $\mathrm{513^3}$ & 4.85 & $10$ & $24$ & $482$ & 11.4 & 11.4 & 11.4 & 0.745 & 2.94 & 140 & 140 \\
			$0.5$ & $\mathrm{1 \times 10^7}$ & $\mathrm{513^3}$ & 3.28 & $8$ & $17$ & $874$ & 15.9 & 16.0 & 16.0 & 0.682 & 2.95 & 91 & 91 \\
			$0.5$ & $\mathrm{3 \times 10^7}$ & $\mathrm{513^3}$ & 2.30 & $7$ & $13$ & $1480$ & 21.6 & 21.8 & 21.6 & 0.550 & 2.73 & 48 & 48\\
			$0.5$ & $\mathrm{1 \times 10^8}$ & $\mathrm{513^3}$ & 1.55 & $5$ & $9$ & $2610$ & 30.6 & 30.8 & 30.6 & 0.475 & 2.58 & 37 & 37 \\ 
			$1$ & $\mathrm{1 \times 10^6}$ & $\mathrm{257^3}$ & 4.92 & $7$ & $17$ & $147$ & 8.18 & 8.45 & 8.48 & 0.765 & 2.83 & 101 & 101\\
			$1$ & $\mathrm{2 \times 10^6}$ & $\mathrm{257^3}$ & 3.94 & $7$ & $14$ & $213$ & 10.1 & 10.1 & 10.2 & 0.791 & 2.98 & 101 & 101\\
			$1$ & $\mathrm{5 \times 10^6}$ & $\mathrm{257^3}$ & 2.90 & $6$ & $11$ & $340$ & 13.3 & 13.3 & 13.4 & 0.709 & 2.97 & 101 & 101 \\
			$1$ & $\mathrm{1 \times 10^7}$ & $\mathrm{257^3}$ & 2.31 & $5$ & $9$ & $491$ & 16.3 & 16.3 & 16.4 & 0.679 & 2.93 & 101 & 101 \\
			$1$ & $\mathrm{2 \times 10^7}$ & $\mathrm{257^3}$ & 1.85 & $5$ & $7$ & $702$ & 19.8 & 19.7 & 19.9 & 0.682 & 2.91 & 91 & 91 \\
			$1$ & $\mathrm{5 \times 10^7}$ & $\mathrm{513^3}$ & 2.73 & $7$ & $11$ & $1100$ & 26.0 & 26.0 & 26.1 & 0.561 & 2.81 & 103 & 103 \\
			$1$ & $\mathrm{1 \times 10^8}$ & $\mathrm{513^3}$ & 2.19 & $6$ & $9$ & $1530$ & 31.4 & 31.3 & 31.5 & 0.512 & 2.69 & 101 & 101 \\
			$1$ & $\mathrm{2 \times 10^8}$ & $\mathrm{513^3}$ &  1.75& $6$ & $8$ & $2170$ & 38.6 & 38.3 & 38.7 & 0.490 & 2.68 & 101 & 101 \\
			$1$ & $\mathrm{5 \times 10^8}$ & $\mathrm{513^3}$ & 1.30 & $5$ & $6$ & $3330$ & 49.2 & 49.6 & 49.2 & 0.437 & 2.51 & 101 & 101 \\
			$1$ & $\mathrm{1 \times 10^9}$ & $\mathrm{1025^3}$ & 2.06 & $7$ & $9$ & $4700$ & 61.2 & 61.6 & 61.4 & 0.426 & 2.35 & 15 & 30 \\
			$1$ & $\mathrm{2 \times 10^9}$ & $\mathrm{1025^3}$ & 1.62 & $7$ & $8$ & $6580$ & 76.8 & 81.1 & 76.7 & 0.392 & 2.47 & 13 & 26 \\
		\end{tabular}
		\label{table:SimDetails}
	\end{ruledtabular}
\end{table*}
\begin{table*}
	\caption{Details of our direct numerical simulations performed in a cubical box for $\mathrm{Pr} > 1$: the Prandtl number (Pr), the Rayleigh Number (Ra),  the grid size, the ratio of the Batchelor length scale~\cite{Batchelor:JFM1959_largeSc} ($\eta_T$) to the mesh width $\Delta x$, the number of grid points in viscous and thermal boundary layers ($N_{\mathrm{VBL}}$ and $N_{\mathrm{TBL}}$ respectively), the Reynolds number (Re), Nusselt number computed using $\langle u_zT \rangle$ and the exact relations given by Eqs.~(\ref{eq:SS_Viscous}) and (\ref{eq:SS_Thermal}) (Nu, $\mathrm{Nu}_u$, and $\mathrm{Nu}_T$ respectively), the ratio of the total viscous dissipation rate in the boundary layer ($\tilde{D}_{u,\mathrm{BL}}$) and that in the bulk ($\tilde{D}_{u,\mathrm{bulk}}$), the ratio of the total thermal dissipation rate in the boundary layer ($\tilde{D}_{T,\mathrm{BL}}$) and that in the bulk ($\tilde{D}_{T,\mathrm{bulk}}$), the number of non-dimensional time units ($t_\mathrm{ND}$) and snapshots over which the quantities are averaged.}
	\begin{ruledtabular}
		\begin{tabular}{c c c c c c c c c c c c c c}
			$\mathrm{Pr}$ & $\mathrm{Ra}$ & Grid size & $\eta_T /\Delta x$ & $N_{\mathrm{VBL}}$ & $N_{\mathrm{TBL}}$ & Re & Nu & $\mathrm{Nu}_u$ & $\mathrm{Nu}_T$ & $\frac{\tilde{D}_{u,\mathrm{BL}}}{\tilde{D}_{u,\mathrm{bulk}}}$ & $\frac{\tilde{D}_{T,\mathrm{BL}}}{\tilde{D}_{T,\mathrm{bulk}}}$ & $t_\mathrm{ND}$ & Snapshots\\
			\hline 
			6.8 & $1 \times 10^6$ & $257^3$ & 5.02 & 9 & 17 & 24.9 & 7.90 & 7.87 & 7.87 & 0.822 & 3.08 & 101 & 101\\
			6.8 & $2 \times 10^6$ & $257^3$ & 4.01 & 8 & 15 & 35.6 & 9.46 & 9.43 & 9.48 & 0.744 & 2.94 & 101 & 101\\
			6.8 & $5 \times 10^6$ & $257^3$ & 2.93 & 7 & 11 & 59.7 & 12.9 & 12.9 & 13.0 & 0.646 & 2.97 & 101 & 101\\
			6.8 & $1 \times 10^7$ & $257^3$ & 2.33 & 6 & 9 & 89.2 & 15.9 & 15.8 & 16.0 & 0.605 & 2.93 & 101 & 101\\
			6.8 & $2 \times 10^7$ & $257^3$ & 1.85 & 6 & 8 & 128 & 19.5 & 19.4 & 19.3 & 0.579 & 2.97 & 107 & 101\\
			6.8 & $5 \times 10^7$ & $257^3$ & 1.37 & 5 & 6 & 217 & 26.1 & 25.7 & 25.9 & 0.588 & 2.99 & 101 & 101\\
			6.8 & $1 \times 10^8$ & $513^3$ & 2.18 & 8 & 9 & 314 & 31.6 & 31.6 & 31.7 & 0.614 & 2.85 & 56 & 56\\
			6.8 & $2 \times 10^8$ & $513^3$ & 1.75 & 7 & 8 & 452 & 38.5 & 37.7 & 39.3 & 0.529 & 2.84 & 26 & 51\\
			6.8 & $5 \times 10^8$ & $513^3$ & 1.29 & 7 & 6 & 729 & 50.5 & 50.4 & 50.8 & 0.521 & 2.83 & 58 & 58\\
			6.8 & $1 \times 10^9$ & $1025^3$ & 2.06 & 11 & 9 & 1070 & 65.7 & 61.9 & 62.0 & 0.518 & 2.69 & 14 & 28\\
			6.8 & $2 \times 10^9$ & $1025^3$ & 1.64 & 10 & 8 & 1520 & 77.0 & 77.6 & 77.5 & 0.463 & 2.83 & 20 & 40\\
			6.8 & $5 \times 10^9$ & $1025^3$ & 1.22 & 9 & 6 & 2400 & 101 & 101 & 101 & 0.440 & 2.72 & 17 & 33\\ 
			50 & $1 \times 10^6$ & $513^3$ & 9.92 & 17 & 33 & 3.53 & 8.17 & 8.16 & 7.99 & 0.815 & 3.23 & 131 & 131\\
			50 & $2 \times 10^6$ & $513^3$ & 7.96 & 16 & 27 & 5.19 & 9.66 & 9.60 & 9.61 & 0.722 & 3.42 & 51 & 51\\
			50 & $5 \times 10^6$ & $513^3$ & 5.74 & 14 & 20 & 9.38 & 13.8 & 13.7 & 13.5 & 0.627 & 3.19 & 130 & 130\\
			50 & $1 \times 10^7$ & $513^3$ & 4.58 & 13 & 17 & 14.0 & 16.7 & 16.7 & 16.2 & 0.581 & 3.12 & 65 & 65 \\
			50 & $2 \times 10^7$ & $513^3$ & 3.67 & 12 & 14 & 21.1 & 20.2 & 20.1 & 20.0 & 0.525 & 3.13 & 55 & 55 \\
			50 & $5 \times 10^7$ & $513^3$ & 2.72 & 11 & 11 & 35.2 & 26.4 & 26.2 & 26.0 & 0.489 & 3.07 & 57 & 57 \\
			50 & $1 \times 10^8$ & $513^3$ & 2.18 & 10 & 9 & 50.8 & 31.8 & 31.6 & 31.6 & 0.436 & 2.92 & 111 & 111 \\
			50 & $2 \times 10^8$ & $513^3$ & 1.74 & 9 & 8 & 76.4 & 38.7 & 38.8 & 38.7 & 0.433 & 3.10 & 101 & 101 \\
			50 & $5 \times 10^8$ & $513^3$ & 1.29 & 9 & 6 & 137 & 51.8 & 51.6 & 50.4 & 0.481 & 2.88 &62 & 62\\
			50 & $1 \times 10^9$ & $513^3$ & 1.03 & 8 & 5 & 202 & 61.5 & 63.0 & 69.3 & 0.599 & 2.79 & 101 & 101 \\ 
			100 & $1 \times 10^6$ & $257^3$ & 5.01 & 10 & 17 & 1.80 & 7.94 & 7.93 & 7.94 & 1.04 & 3.41 & 259 & 259\\
			100 & $2 \times 10^6$ & $257^3$ & 3.91 & 9 & 14 & 2.78 & 10.4 & 10.3 & 10.2 & 0.862 & 3.42 & 263 & 263\\
			100 & $5 \times 10^6$ & $257^3$ & 2.87 & 8 & 10 & 4.90 & 13.9 & 13.9 & 14.0 & 0.731 & 3.36 & 153 & 153\\
			100 & $1 \times 10^7$ & $257^3$ & 2.30 & 7 & 9 & 7.02 & 16.8 & 16.7 & 16.6 & 0.585 & 3.30 & 101 & 101\\
			100 & $2 \times 10^7$ & $257^3$ & 1.84 & 7 & 7 & 9.91 & 20.1 & 20.0 & 19.9 & 0.485 & 3.00 & 101 & 101\\
			100 & $5 \times 10^7$ & $257^3$ & 1.37 & 6 & 6 & 17.1 & 26.1 & 25.9 & 26.1 & 0.467 & 3.20 & 101 & 101\\
			100 & $1 \times 10^8$ & $513^3$ & 2.18 & 10 & 9 & 26.0 & 31.8 & 31.7 & 31.7 & 0.433 & 2.96 & 107 & 107\\
			100 & $2 \times 10^8$ & $513^3$ & 1.74 & 9 & 8 & 37.5 & 39.1 & 38.8 & 38.8 & 0.373 & 3.08 & 108 & 108\\
			100 & $5 \times 10^8$ & $513^3$ & 1.30 & 10 & 6 & 71.4 & 49.7 & 49.2 & 50.3 & 0.429 & 2.95 & 86 & 86
		\end{tabular}
		\label{table:SimDetails_Pr>1}
	\end{ruledtabular}
\end{table*}
We perform  direct numerical simulations of RBC by solving Eqs.~(\ref{eq:NDMomentum})-(\ref{eq:NDContinuity}) in a cubical box of unit dimension using the finite difference code SARAS.~\cite{Verma:SNC2020,Samuel:JOSS2020} We carry out 60 runs   for Pr ranging from 0.02 to 100 and Ra ranging from $5 \times 10^5$ to $5 \times 10^9$. The grid size {\color{black}was} varied from $257^3$ to $1025^3$ depending on parameters. Refer to Tables~\ref{table:SimDetails} and \ref{table:SimDetails_Pr>1} for the simulation details.  


We impose isothermal boundary conditions on the horizontal walls and adiabatic boundary conditions on the sidewalls. No-slip boundary conditions were imposed on all the walls. A second-order Crank-Nicholson scheme was used for time-advancement, with the maximum Courant number kept at 0.2. The solver uses a multigrid method for solving the pressure-Poisson equations. We ensure a minimum of 5 points in the viscous and the thermal boundary layers (see Tables~\ref{table:SimDetails} and \ref{table:SimDetails_Pr>1}); this satisfies the resolution criterion of \citet{Grotzbach:JCP1983}, and \citet{Verzicco:JFM2003}. The simulations are run up to {\color{black}3} to 263 non-dimensional time units ($t_\mathrm{ND}$) after attaining a steady state. For post-processing, we employ central difference method for spatial differentiation and Simpson's method for computing the volume average. 

In order  to resolve the smallest scales of the flow, we ensure that the grid spacing $\Delta x$ is smaller than the Kolmogorov length scale $\eta = (\nu^3 \epsilon_u^{-1})^{1/4}$ for $\mathrm{Pr} \leq 1$ and the Batchelor length scale $\eta_T = (\nu \kappa^2 \epsilon_u^{-1})^{1/4}$ for $\mathrm{Pr}>1$. We numerically compute $\epsilon_u$ and $\epsilon_T$ and use these values to compute $\mathrm{Nu}_u$ and $\mathrm{Nu}_T$ employing Shraimann and Siggia's exact relations~\cite{Shraiman:PRA1990}   [see Eqs.~(\ref{eq:SS_Viscous}) and (\ref{eq:SS_Thermal})]. The Nusselt numbers computed using $\langle u_z T \rangle$ match with $\mathrm{Nu}_u$ and $\mathrm{Nu}_T$ within two percent on an average; this further confirms that our runs are well-resolved (see  Tables~\ref{table:SimDetails} and \ref{table:SimDetails_Pr>1}).  All the above quantities are averaged over 12 to 259 snapshots taken at equal time intervals after attaining a steady state. 


In the next section, {\color{black}we analyse our numerical results, construct the cubic polynomial relation for Re and Nu using the data from our simulations, and compare our revised predictions with those of the GL model.}

\section{Results} \label{sec:Results}

Using our numerical data, we determine the scaling of dissipation rates, boundary layer thicknesses, and the functional forms of $ f_i $.  We {\color{black} construct the relations for Re and Nu given by Eqs.~(\ref{eq:Equation1}) and (\ref{eq:Equation2})  using these inputs and compare the revised predictions with those of the original GL model.}
We also analyse how {\color{black} the proposed relation} performs in the limit of $\tilde{D}_{u,\mathrm{bulk}} \gg \tilde{D}_{u,\mathrm{BL}}$ and vice-versa.

\subsection{Viscous and thermal dissipation rates} \label{subsec:EuEt}
\begin{figure}[t]
	\includegraphics[scale=0.375]{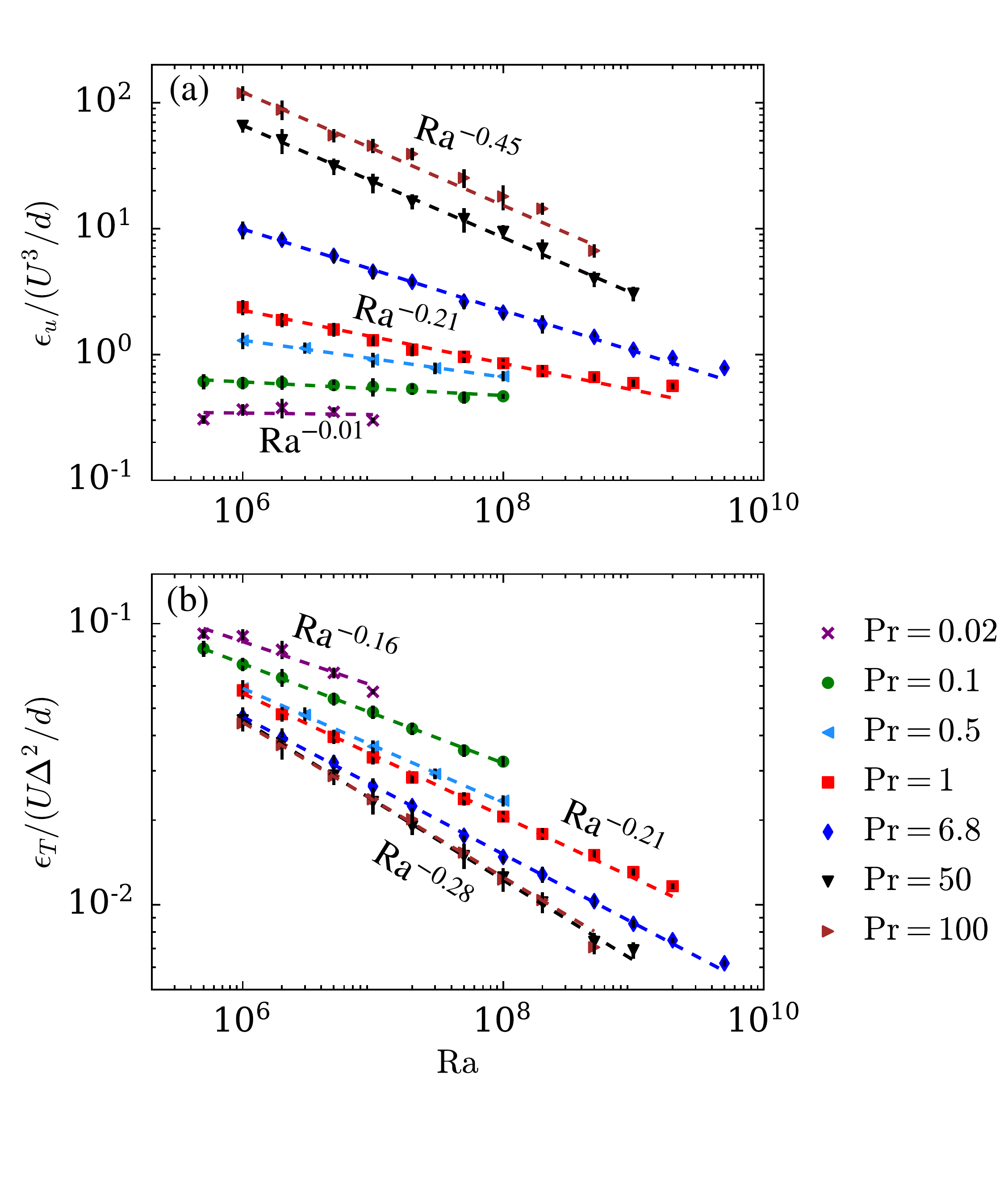}
	\caption{(color online) 
	Plots of (a): normalized viscous dissipation rate vs. Ra, (b): normalized thermal dissipation rate vs {\color{black}Ra}. The error bars represent the standard deviation of the dataset with respect to the temporal average. Both the viscous and thermal dissipation rates exhibit additional Ra dependence.}
	\label{fig:Dissipation}
\end{figure}
Here, we examine the scaling of viscous and thermal dissipation rates and explore how their scaling deviates from that of free turbulence. First, we present theoretical arguments on the above scaling, following which we verify our arguments with our numerical results.
	
{\color{black}In free turbulence, the viscous and scalar dissipation rates are estimated as follows:
\begin{equation}
\epsilon_u \sim \frac{U^3}{d}, \quad \epsilon_T \sim \frac{U \Delta^2}{d}.
\label{eq:Eu_Et}
\end{equation}
However, in wall-bounded convection, the scaling of the dissipation rates is different. To understand this, let us rewrite the exact relations of \citet{Shraiman:PRA1990} given by Eqs.~(\ref{eq:SS_Viscous}) and (\ref{eq:SS_Thermal}) as  
\begin{eqnarray}
\epsilon_u &=& \frac{U^3}{d}\frac{1}{\mathrm{Re}^3} (\mathrm{Nu} - 1) \frac{\mathrm{Ra}}{\mathrm{Pr}^2} , \label{eq:SS_Viscous_modified} \\
\epsilon_T &=& \frac{U \Delta^2}{d} \frac{1}{\mathrm{RePr}} \mathrm{Nu}. \label{eq:SS_Thermal_modified}
\end{eqnarray} 
Recall from Sec.~\ref{sec:Introduction} that the Reynolds number scales as $\mathrm{Re} \sim \mathrm{Ra}^{1/2}$ for $\mathrm{Pr} \sim 1$ and $\mathrm{Re} \sim \mathrm{Ra}^{0.6}$ for $\mathrm{Pr} \gg 1$, and the Nusselt number scales as $\mathrm{Nu} \sim \mathrm{Ra}^{0.3}$ for $\mathrm{Pr} \gtrsim 1$. Substitution the above relations in Eqs.~(\ref{eq:SS_Viscous_modified}) and (\ref{eq:SS_Thermal_modified}) yields
\begin{equation}
\epsilon_u \sim 
\begin{cases}
\frac{U^3}{d} \mathrm{Ra}^{-0.2}, \quad \mathrm{Pr} \sim 1, \\
\frac{U^3}{d} \mathrm{Ra}^{-0.5}, \quad \mathrm{Pr} \gg 1,
\end{cases}
\label{eq:Eu_Pr_geq_1}
\end{equation}
instead of $U^3/d$, and 
\begin{equation}
\epsilon_T \sim 
\begin{cases}
\frac{U\Delta^2}{d} \mathrm{Ra}^{-0.2}, \quad \mathrm{Pr}\sim 1, \\
\frac{U \Delta^2}{d} \mathrm{Ra}^{-0.3} \quad \mathrm{Pr} \gg 1.
\end{cases}
\label{eq:Et_Pr_geq_1}
\end{equation} 
instead of $U \Delta^2/d$. \citet{Pandey:PF2016} and \citet{Pandey:PRE2016} argue that the additional Ra dependence is due to the suppression of nonlinear interactions due to the presence of walls. Some Fourier modes that are otherwise present in free turbulence are absent in wall-bounded RBC; this results in several channels of nonlinear interactions and energy cascades to be blocked~\cite{Verma:book:BDF}. Note that the horizontal walls  seem to have a more pronounced effect on the aforementioned suppression than the lateral walls, as \citet{Schmidt:JFM2012} observed passive scalar scaling for homogeneous laterally confined RBC. In addition, buoyancy also appears to suppress the energy cascade rate,~\cite{Bhattacharya:PF2019b} similar to the role played by magnetic field in magnetohydrodynamic turbulence.~\cite{Verma:PP2020}

Now, for $\mathrm{Pr} \ll 1$, recall that $\mathrm{Re} \sim \mathrm{Ra}^{0.42}$ and $\mathrm{Nu} \sim \mathrm{Ra}^{0.25}$ (see Sec.~\ref{sec:Introduction}). Substitution of these expressions in Eqs.~(\ref{eq:SS_Viscous_modified}) and (\ref{eq:SS_Thermal_modified}) yields
\begin{equation}
\epsilon_u \sim \frac{U^3}{d}, \quad \epsilon_T \sim \frac{U \Delta^2}{d} \mathrm{Ra}^{-0.17}. \label{eq:Eu_Et_Pr_ll_1}
\end{equation} 
Thus, the viscous dissipation rate scales similar to free turbulence for small Pr. However, the additional Ra dependence is still present in the scaling of thermal dissipation rates because of the presence of thick thermal boundary layers.}

Using our data, we numerically compute the viscous and thermal dissipation rates and normalize them with $U^3/d$ and $U \Delta^2/d$ respectively. We plot the normalized dissipation rates versus Ra and exhibit these plots in Figs~\ref{fig:Dissipation}(a,b). We observe that for small Pr, the normalized viscous dissipation rate is independent of Ra, whereas for larger Pr, the aforementioned quantity decreases with Ra. The decrease becomes steeper as Pr increases, with $\epsilon_u/(U^3d^{-1}) \sim \mathrm{Ra}^{-0.21}$ for $\mathrm{Pr}=1$ and $\sim \mathrm{Ra}^{-0.45}$ for $\mathrm{Pr}=100$. The normalized thermal dissipation rate decreases with Ra for all Pr, with $\epsilon_T/(U \Delta^2 d^{-1}) \sim \mathrm{Ra}^{-0.15}$ for $\mathrm{Pr}=0.02$ to $ \sim \mathrm{Ra}^{-0.28}$ for $\mathrm{Pr}=100$, which are consistent with the earlier estimates.

In the next subsection, we discuss the computations of the boundary layer thicknesses and their dependence on Re and Nu for different Pr.
 
\subsection{Boundary layer thicknesses} \label{subsec:BLThickness}
\begin{figure}[t]
	\includegraphics[scale=0.4]{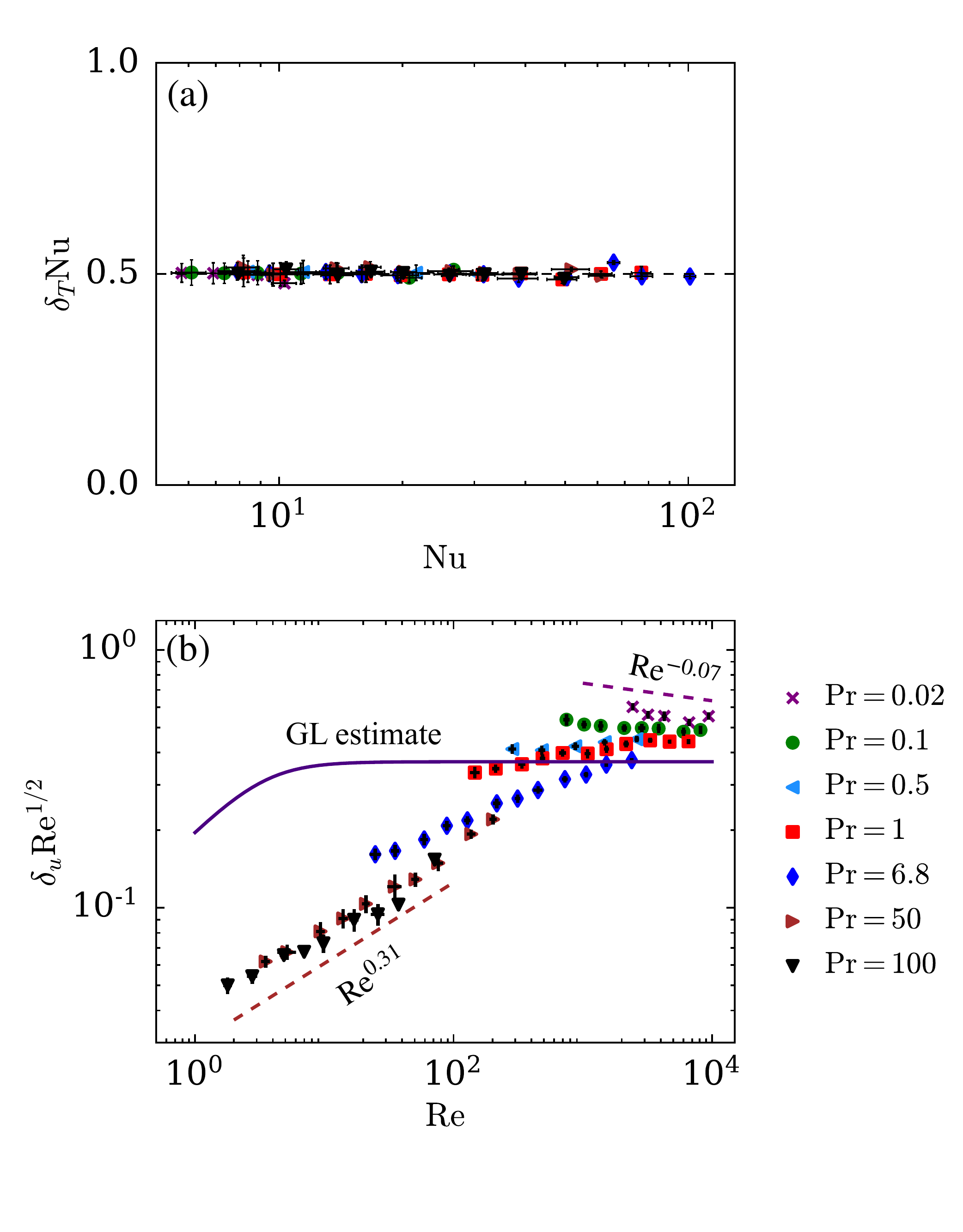}
	\caption{(color online) Plots of (a): normalized thermal boundary layer thickness vs. Nu, (b): normalized viscous boundary layer thickness vs Re. The error bars represent the standard deviation of the dataset with respect to the temporal average. The viscous boundary layer thickness deviates from the Prandtl-Blassius relation of $\delta_u \sim \mathrm{Re^{-1/2}}$, as well as from Grossmann and Lohse's estimate of $g(\sqrt{\mathrm{Re}_c/\mathrm{Re}})$.}
	\label{fig:BLThickness}
\end{figure}
There are several ways to define the viscous and the thermal boundary layer thicknesses in RBC.~\cite{Ahlers:RMP2009,Scheel:JFM2012}
In our paper, the viscous boundary layer thickness $\delta_u$ is defined as the depth where a linear fit of the velocity profile near the wall intersects with the tangent to the velocity profile at its local maximum. Similarly, the thermal boundary layer thickness $\delta_T$ is defined as the depth where a linear fit of the temperature profile near the wall intersects with the mean temperature $T=0.5$. 
The above methods are described in detail in Refs.~\cite{Breuer:PRE2004,Ahlers:RMP2009,Scheel:JFM2012}.

Using the data generated from our simulations, we first compute the thicknesses of the thermal and viscous boundary layers. We report the average thicknesses of the viscous boundary layers near all the six walls and the thermal boundary layers near the top and bottom walls. We examine the validity of the Prandtl-Blasius relation of $\delta_u \sim \mathrm{Re}^{-0.5}$ for the viscous boundary layers and $\delta_T = 0.5\mathrm{Nu}^{-1}$ for the thermal boundary layers. Towards this objective, we plot $\delta_T \mathrm{Nu}$ versus Nu in Fig~\ref{fig:BLThickness}(a) and  $\delta_u \mathrm{Re}^{1/2}$ versus Re in Fig~\ref{fig:BLThickness}(b). 
 
We observe from Fig~\ref{fig:BLThickness}(a) that $\delta_T \mathrm{Nu} \approx 1/2$, independent of Nu, which is consistent with the definition.  On the other hand, from Fig~\ref{fig:BLThickness}(b), it is evident that $\delta_u \mathrm{Re}^{1/2}$ is constant  in Re only for $\mathrm{Pr}=0.5$ and 0.1.  However, $\delta_u \mathrm{Re}^{1/2}$  increases   as $\sim \mathrm{Re^{0.31}}$ for large Pr; and  decreases marginally as $\sim \mathrm{Re^{-0.07}}$ for $\mathrm{Pr}=0.02$.  This shows that for large Pr, $\delta_u$ becomes a weak function of Re; this is consistent with the observation of \citet{Breuer:PRE2004}  We also plot Grossmann and Lohse's \cite{Grossmann:PRL2001} estimate of viscous boundary layer thickness which is given by $g(\sqrt{\mathrm{Re}_c/\mathrm{Re}})$; here $g(x) = x(1+x^4)^{-1/4}$ and $\mathrm{Re}_c=3.401$. It is clear that the Grossmann and Lohse's estimate deviates significantly from the actual values.

Therefore, we cannot assume $\delta_u \sim g(\mathrm{ Re}^{-1/2})$ for viscous boundary layers in RBC, and it is more prudent to obtain the scaling of $f_2 \delta_u^{-1}$ with Ra, where $f_2$ is the function from Eq.~(\ref{eq:fnuU^2/d^2}).  The above deviation from Prandtl-Blasius profile has also been observed in previous studies.~\cite{Scheel:JFM2012,Shi:JFM2012,Bhattacharya:PF2018} {\color{black}This is because $\delta_u \sim \mathrm{Re}^{-1/2}$ is valid asymptotically for very large Reynolds numbers.}~\cite{Landau:book:Fluid}

\subsection{$f_i$ versus Ra for different Pr} \label{subsec:Coefficients}
\begin{figure*}[t]
	\includegraphics[scale=0.4]{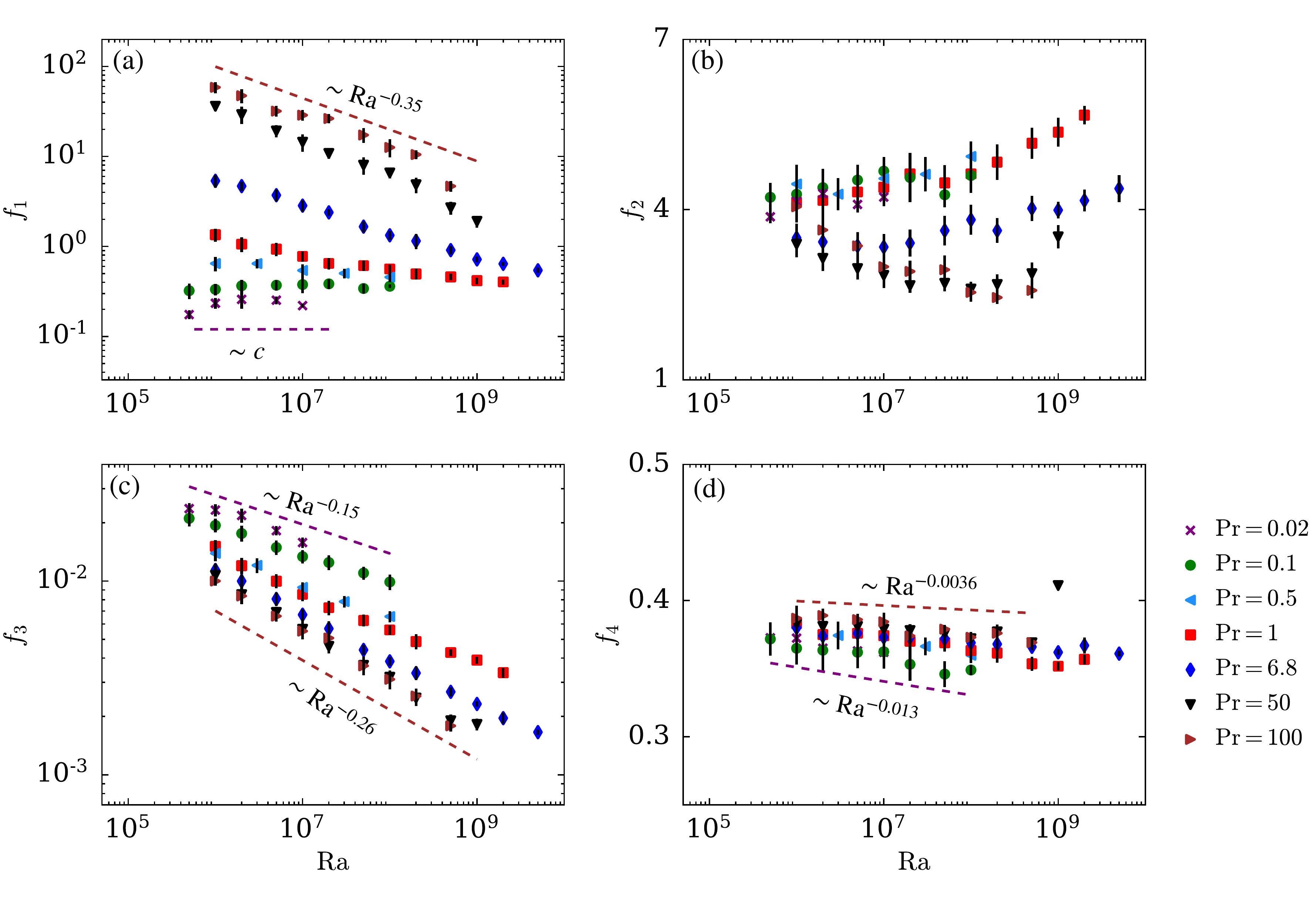}
	\caption{(color online) Plots of {\color{black}(a)} $f_1$, {\color{black}(b)} $f_2$, {\color{black}(c)} $f_3$, and {\color{black}(d)} $f_4$ vs. Ra. The error bars represent the standard deviation of the dataset with respect to the temporal average. $f_2$ remains roughly independent of Ra and Pr albeit with fluctuations; however, $f_1$, $f_3$ and $f_4$ decrease with Ra. }
	\label{fig:Coefficients}
\end{figure*}

In this subsection, we numerically compute $f_i$ using our simulation data and discuss how these quantities vary with Ra for different Pr. We also obtain the limiting cases for the scaling of $f_i$ with Ra. 

We numerically compute the total viscous and thermal dissipation rates in the bulk and in the boundary layers for all the simulation runs. 
Using these values and boundary layer thicknesses, we compute $f_1$, $f_2$, $f_3$, and $f_4$ and plot them versus Ra  in Fig.~\ref{fig:Coefficients}. We observe that $f_1$ and $f_3$ are, in general, not constants as in free turbulence. 
$f_1$ decreases with Ra except for $\mathrm{Pr}=0.1$ and $0.02$, where it is nearly constant. The above decrease is more prominent for large Pr ($\geq 50$), where $f_1 \sim \mathrm{Ra}^{-0.35}$. In a similar fashion, $f_3$ also decreases with Ra for all Pr, and is more pronounced for large Pr ($f_3 \sim \mathrm{Ra}^{-0.26}$) and less pronounced for small Pr ($f_3 \sim \mathrm{Ra}^{-0.15}$). 
{\color{black}The above observations imply that the scaling of the dissipation rates in the bulk is similar to that in the entire volume~\cite{Bhattacharya:PF2018,Bhattacharya:PF2019}~(see Section~\ref{subsec:EuEt}). This is because the bulk occupies a large fraction of the total volume and its contribution to the total dissipation is significant.}~\cite{Bhattacharya:PF2018,Bhattacharya:PF2019} 

The Ra and Pr dependence of $f_2$ cannot be clearly established from Fig.~\ref{fig:Coefficients}(b); we can only infer that $f_2$ is independent of Ra and Pr, albeit with significant fluctuations. This is consistent with $\epsilon_{u,\mathrm{BL}} \sim \nu U^2/\delta_u^2$ as predicted by \citet{Grossmann:JFM2000,Grossmann:PRL2001}.   The function $f_4$ of Fig.~\ref{fig:Coefficients}(d) appears flat, but a careful examination shows that $f_4$ decreases  weakly with Ra, with  $f_4 \sim \mathrm{Ra}^{-0.013}$ for small Pr and  $f_4 \sim \mathrm{Ra}^{-0.0036}$ for large Pr. The reason for the marginal decrease of $ f_4 $ with Ra needs investigation and is not in the scope of this paper.

\begin{figure}[t]
	\includegraphics[scale=0.13]{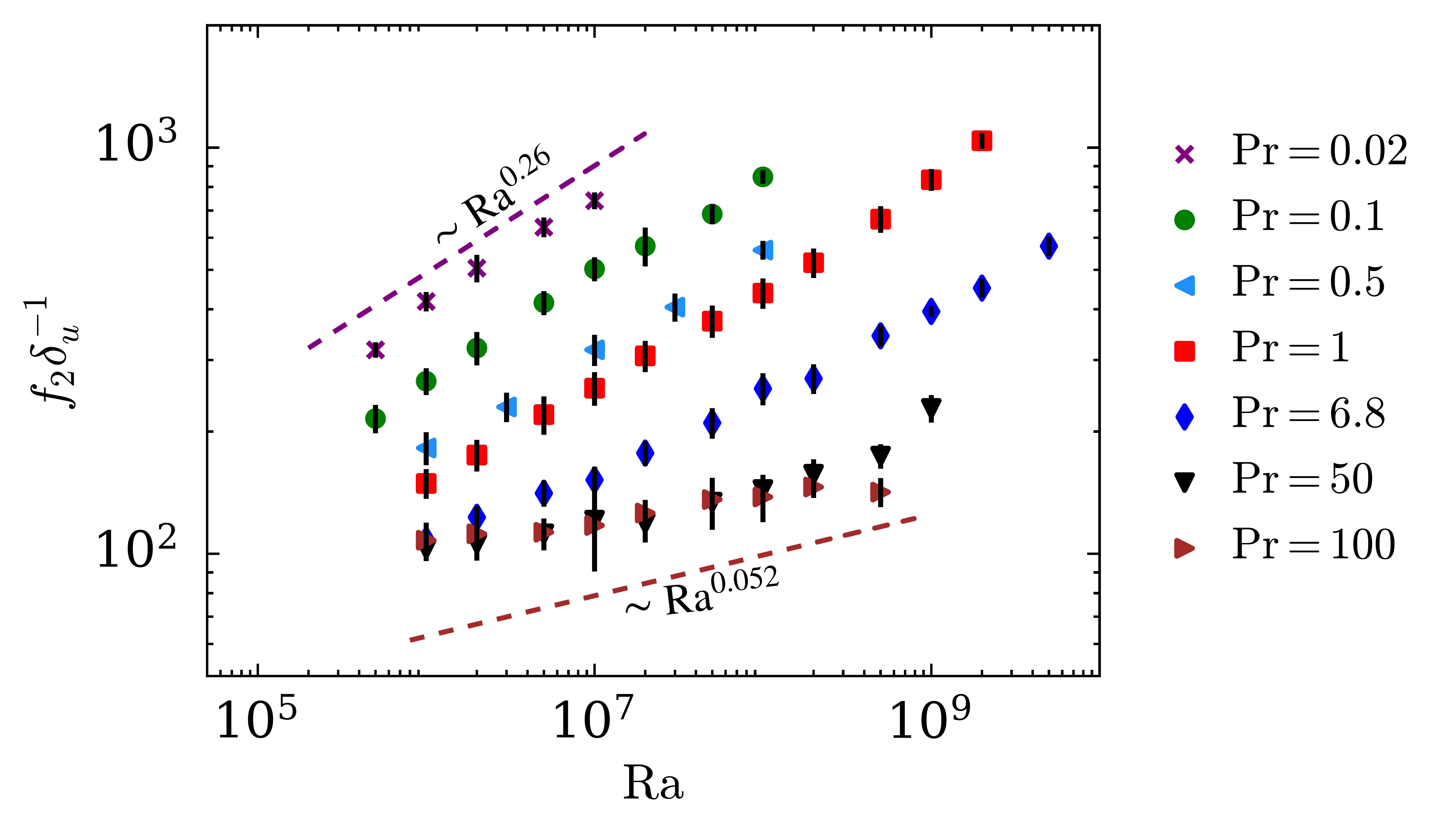}
	\caption{(color online) Plot of $f_2 \delta_u^{-1}$ vs Ra. The error bars represent the standard deviation of the dataset with respect to the temporal average. The dependence of $f_2 \delta_u^{-1}$ on Ra is stronger for small Pr and becomes weaker as Pr increases.}
	\label{fig:c2p}
\end{figure}

As discussed earlier, the solution of Eq.~(\ref{eq:Cubic_Re}) for Re and Nu depends on the quantity $f_2 \delta_u^{-1}$.   Hence, we plot this quantity versus Ra for different Pr in Fig.~\ref{fig:c2p}.  Since $f_2$ is nearly constant, $f_2\delta_u^{-1}$ is inversely proportional to the viscous boundary layer thickness. Thus, $f_2 \delta_u^{-1}$ increases marginally for large Pr ($\sim \mathrm{Ra}^{0.052}$) and steeply for small Pr ($\sim \mathrm{Ra}^{0.26}$), which is in agreement with the scaling of viscous boundary layer thickness discussed in Sec.~\ref{subsec:BLThickness}.

In the next subsection, we describe the machine-learning tools used to determine the functional forms of $f_i$.


\subsection{Machine learning algorithm to obtain $f_i(\mathrm{Ra,Pr})$}
\label{subsec:ML}
So far, we have examined the variation of $f_i$ with only Ra for different Prandtl numbers and obtained the limiting cases. Now, using machine learning and matching functions, we will combine these scalings to determine $f_i$ as functions of both Ra and Pr. We make use of the machine-learning software WEKA~\cite{Weka:2009} for obtaining the functional forms of $f_i$. The values of $f_i$ computed for every Ra and Pr using our simulation data serve as training sets for our machine learning algorithm. For simplicity,  we will look for a power-law relation of the form $f_i = A\mathrm{Ra}^\alpha \mathrm{Pr}^\beta$, take logarithms of this expression, and employ \textit{linear regression} to obtain $A$, $\alpha$, and $\beta$. The linear regression algorithm works by estimating coefficients for a hyperplane that best fits the training data using least squares method.

\begin{figure}[t]
	\includegraphics[scale=0.39]{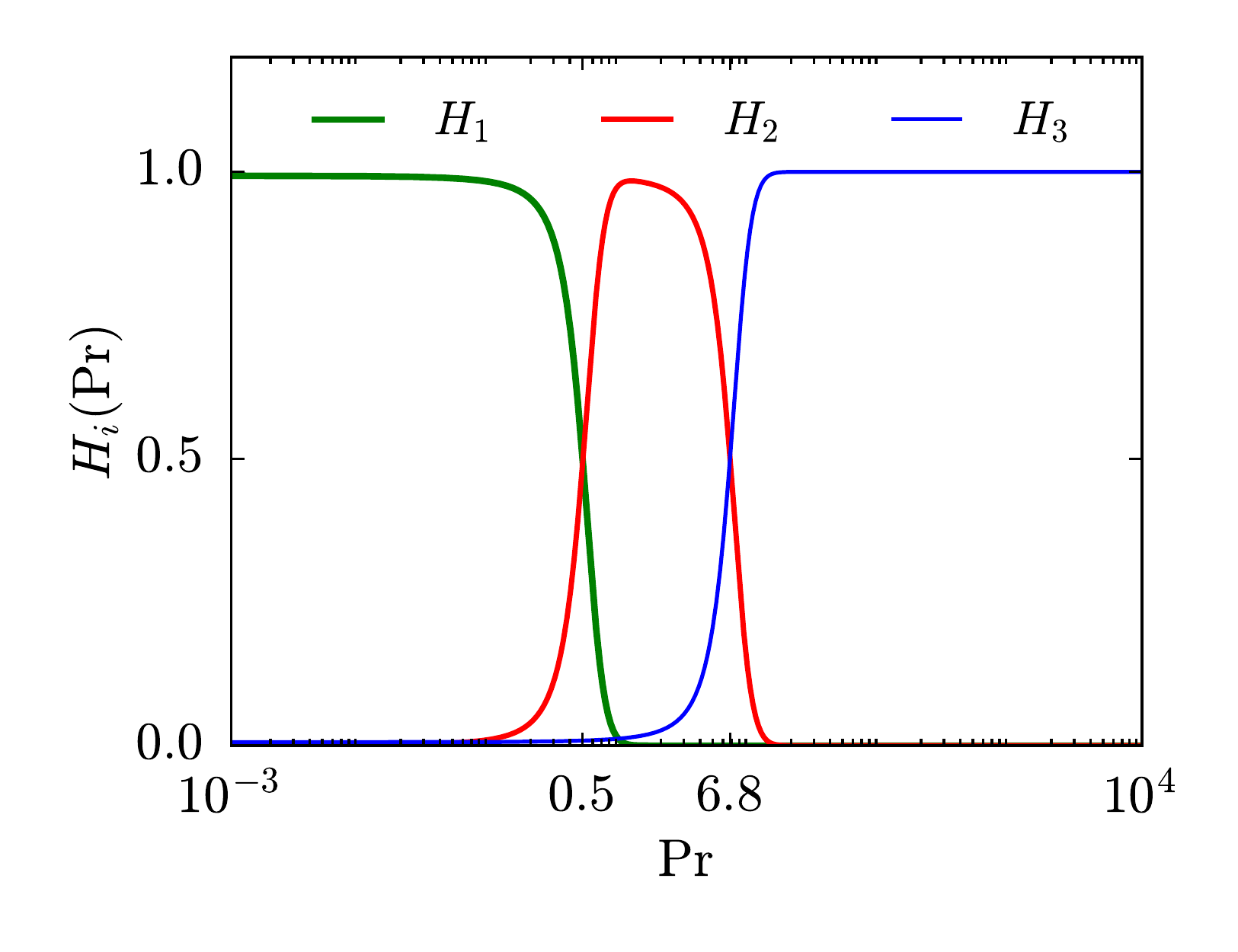}
	\caption{(color online) Plot of the matching functions $H_i(\mathrm{Pr})$ vs. Pr. $H_1$, $H_2$, and $H_3$ become unity in the regimes given by $\mathrm{Pr}<0.5$, $0.5<\mathrm{Pr}<6.8$, and $\mathrm{Pr}>6.8$ respectively. They attain the value of $1/2$ at the regime boundaries and become negligible outside their respective regimes.}
	\label{fig:Match}
\end{figure}

Since the dependence of $f_i$ on Ra is not uniform~(see Sec.~\ref{subsec:Coefficients}), we split our parameter space into three regimes such that for each regime, the scaling of $f_i$ with Ra is approximately the same. We choose the regimes as follows:
\begin{eqnarray}
\mbox{Small Pr}&:& \quad \mathrm{Pr} \leq 0.5, \nonumber \\
\mbox{Moderate Pr}&:& \quad 0.5 \leq \mathrm{Pr} \leq 6.8, \nonumber \\
\mbox{Large Pr}&:& \quad \mathrm{Pr} \geq 6.8. \nonumber
\end{eqnarray} 
We then determine the prefactor $A$ and the exponents $\alpha$ and $\beta$ for each regime. 
To ensure continuity between the regimes, we introduce the following matching functions: 
\begin{eqnarray}
H_1(\mathrm{Pr}) &=& \frac{1}{1+e^{-k_{1}(0.5-\mathrm{Pr})}}, \label{eq:H1} \\
H_2(\mathrm{Pr}) &=& \frac{1}{1+e^{-k_{1}(\mathrm{Pr}-0.5)}} - \frac{1}{1+e^{-k_{2}(\mathrm{Pr}-6.8)}}, \label{eq:H2} \\
H_3(\mathrm{Pr}) &=& \frac{1}{1+e^{-k_{2}(\mathrm{Pr}-6.8)}}, \label{eq:H3} 
\end{eqnarray} 
where $k_1$ and $k_2$ are taken to be 10 and 0.75 respectively. The functions $H_1$, $H_2$, and $H_3$ become unity inside the regimes given by $\mathrm{Pr}<0.5$, $0.5<\mathrm{Pr}<6.8$, and $\mathrm{Pr}>6.8$ respectively, and become negligible outside their regimes. The value of these functions is 1/2 in the boundaries of their respective regimes. See Fig.~\ref{fig:Match} for an illustration of the behavior of the matching functions.
Using these functions and employing regression for each regime, we obtain the following fits for $f_i$:
\begin{eqnarray}
f_1 &=& 0.67 H_1\mathrm{Pr}^{0.28} + 27 H_2 \mathrm{Ra}^{-0.21}\mathrm{Pr}^{0.55} \nonumber \\ 
&& + 170 H_3 \mathrm{Ra}^{-0.34}\mathrm{Pr}^{0.78}, \label{eq:c1_Fit} \\
\frac{f_2}{\delta_u} &=& 4.4 H_1 \mathrm{Ra}^{0.25}\mathrm{Pr}^{-0.26} + 7.4 H_2 \mathrm{Ra}^{0.22}\mathrm{Pr}^{-0.29}  \nonumber \\
&&+ 27 H_3 \mathrm{Ra}^{0.14}\mathrm{Pr}^{-0.18}, \label{eq:c2p_Fit} \\
f_3 &=& 0.095 H_1\mathrm{Ra}^{-0.15}\mathrm{Pr}^{-0.17} 
 {\color{black}+} 0.25 H_2\mathrm{Ra}^{-0.21}\mathrm{Pr}^{-0.17} \nonumber \\
&&+ 0.45 H_3\mathrm{Ra}^{-0.25}\mathrm{Pr}^{-0.093}, \label{eq:c3_Fit} \\
f_4 &=& 0.46 H_1\mathrm{Ra}^{-0.013}\mathrm{Pr}^{0.010} 
+ 0.43 H_2\mathrm{Ra}^{\color{black}{-0.0081}}\mathrm{Pr}^{\color{black}{0.0053}} \nonumber \\
&&+ 0.39 H_3\mathrm{Ra}^{-0.0036}\mathrm{Pr}^{0.0093}. \label{eq:c4_Fit}
\end{eqnarray}

{\color{black}
	The average deviation between the $f_i$'s predicted by the fits and the actual values are 24\%, 19\%, 12\%, and 58\% for $f_1$, $f_2/\delta_u$, $f_3$, and $f_4$ respectively.} As we will see later, {\color{black}incorporation of the aforementioned functional forms results in more accurate predictions than the GL model}; thus the above uncertainty in $f_i$ is acceptable. {\color{black}In Appendix~\ref{sec:Robustness}, we employ the same regression algorithm over a reduced training set consisting of half of our data-points and show that the fits so obtained are close to Eqs.~(\ref{eq:c1_Fit}-\ref{eq:c4_Fit}). Thus, the estimated parameter values are reasonably robust.
} 

Having obtained the functional forms of $f_i$, we can plug them in Eqs.~(\ref{eq:Cubic_Re}) and (\ref{eq:Nu_RePr}) to complete the {\color{black}relation for Re and Nu.}   
We  remark that $f_i$ obtained above are valid for RBC cells with unit aspect ratio. We suspect that they are weak functions of aspect ratio; this study will be taken up  in  future work. Further, efforts are ongoing to make the functional forms of $f_i(\mathrm{Ra,Pr})$ more compact.

\subsection{\color{black}Enhancement of the GL model} \label{subsec:Comparison}
In this subsection, {\color{black}we will examine the enhancement of the GL model brought about by using the obtained functional forms for the prefactors of the dissipation rates.
We will test both, the original GL model and the revised estimates} with our numerical results, as well as those of \citet{Scheel:PRF2017} ($\mathrm{Pr} =0.005$ and $0.02$), \citet{Wagner:PF2013} ($\mathrm{Pr}=0.7$), \citet{Emran:JFM2008} ($\mathrm{Pr}=0.7$),  \citet{Kaczorowski:JFM2013} ($\mathrm{Pr}=4.38$), and \citet{Horn:JFM2013} ($\mathrm{Pr}=2547.9$). We also include the experimental results of \citet{Cioni:JFM1997} ($\mathrm{Pr}=0.02$), and \citet{Niemela:JFM2001} ($\mathrm{Pr}=0.7$) for our comparisons. The simulations of \citet{Wagner:PF2013} and \citet{Kaczorowski:JFM2013} {\color{black}involved} a cubical cell like ours, whereas the rest of the above simulations and experiments involved a cylindrical cell. All  the above work involve RBC cells with unit aspect ratio.
We compute the percentage deviations ($\mathcal{D}_\mathrm{Re}$ and $\mathcal{D}_\mathrm{Nu}$) between the {\color{black}estimated} and actual values according to the following formula: 	
\begin{equation}
\mathcal{D} = \left | \frac{\mbox{Predicted value} - \mbox{Actual value}}{\mbox{Actual value}} \right | \times 100.
\label{eq:Delta}
\end{equation}
In Table~\ref{table:Model_Comparison}, we list the average of the deviations computed for all the points for every Pr.

\begin{table*}
	\caption{Quantitative comparison between the predictions the GL model and {\color{black}the revised estimates of Nu and Re} for different sets of simulation and experimental data. $\mathcal{D}_\mathrm{Re}$ is the percentage difference between the observed and predicted values of Re, and $\mathcal{D}_\mathrm{Nu}$ is the percentage difference between the observed and predicted values of Nu~[see Eq.~(\ref{eq:Delta})]. Note that no data on Re is available for $\mathrm{Pr=4.38}$.~\cite{Kaczorowski:JFM2013}}
	\begin{ruledtabular}
		\begin{tabular}{c c c c c c c}
			$\mathrm{Pr}$ & Range of Ra   & $\mathcal{D}_\mathrm{Re}$  & $\mathcal{D}_\mathrm{Re}$  & Range of Ra  & $\mathcal{D}_\mathrm{Nu}$  & $\mathcal{D}_\mathrm{Nu}$  \\
			& (Re) & {\color{black}(Revised estimate)} & {\color{black}(GL Model)} & (Nu) & {\color{black}(Revised estimate)} & {\color{black}(GL Model)} \\
			\hline 
			0.005 & $3 \times 10^5$ to $10^7$ & $11\%$ & $48\%$ & $3 \times 10^5$ to $10^7$  & $9.6\%$ & $17\%$ \\
			0.02 & $3 \times 10^5$ to $3 \times 10^9$ & {\color{black}$9.1\%$} & $62\%$ & $3 \times 10^5$ to $3 \times 10^9$ & $10\%$ & $15\%$ \\
			0.1 & $5 \times 10^5$ to $10^8$ & {\color{black}$1.3\%$} & $30\%$ & $5 \times 10^5$ to $10^8$ & $3.1\%$ & $5.0\%$ \\
			0.5 & $10^6$ to $10^8$ & {\color{black}$1.9\%$} & $14\%$ & $10^6$ to $10^8$ & {\color{black} $1.4\%$} & $5.4\%$ \\
			0.7 & $10^5$ to $10^{13}$ & {\color{black}$6.8\%$} & $25\%$ & $10^5$ to $10^{9}$ & {\color{black} $3.8\%$} & $9.9\%$ \\
			1.0 & $10^6$ to $2\times 10^9$ & {\color{black}$2.8\%$} & $20\%$ & $10^6$ to $2\times 10^9$ & {\color{black}$3.6 \%$} & $5.8 \%$ \\
			4.38 & --- & --- & --- & $10^6$ to $3 \times 10^9$ & {\color{black}$5.7 \%$} & $6.3 \%$ \\ 
			6.8 & $10^6$ to $5 \times 10^9$ & {\color{black}$3.4\%$} & $27\%$ & $10^6$ to $5 \times 10^9$ & {\color{black}$5.6\%$} & $6.5\%$ \\
			50 & $10^6$ to $10^9$ & $6.0\%$ & $84\%$ & $10^6$ to $10^9$ & $3.2\%$ & $7.2\%$ \\
			100 & $10^6$ to $5 \times 10^8$ & $3.4 \%$ & $150\%$ & $10^6$ to $5 \times 10^8$ & $2.7\%$ & $3.9\%$ \\
			2547.9 & $10^5$ to $10^9$ & $85\%$ & $560\%$ & $10^5$ to $10^9$ &  $2.3\%$ & $17\%$
		\end{tabular}
		\label{table:Model_Comparison}
	\end{ruledtabular}
\end{table*} 

\begin{figure}[b]
	\includegraphics[scale=0.39]{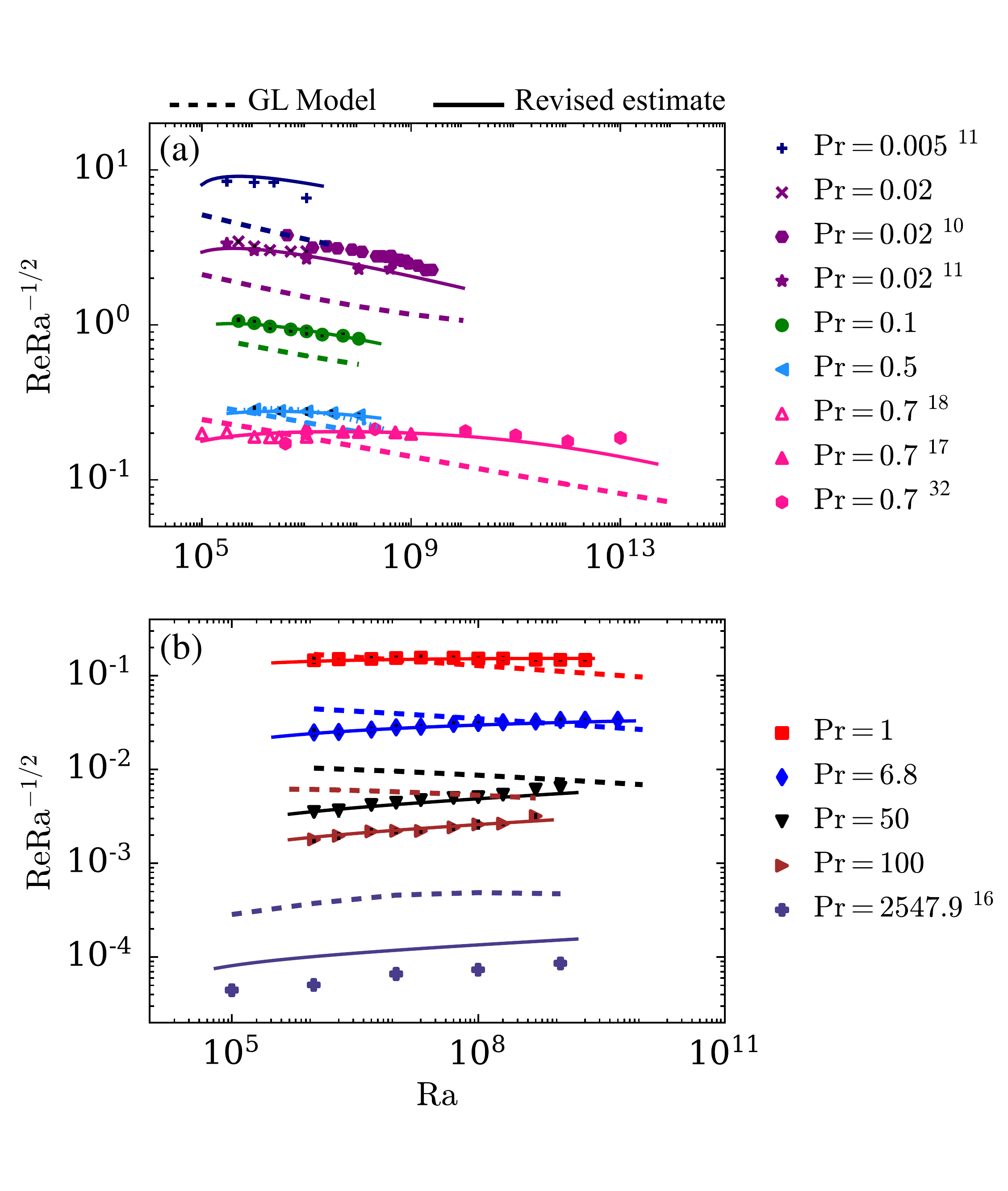}
	\caption{(color online) Comparison between the predictions of Re vs. Ra using {\color{black}the original GL model (dashed curves) and our proposed modifications (solid curves)} with the results from our work and from the literature~\cite{Scheel:PRF2017,Cioni:JFM1997,Emran:JFM2008,Wagner:PF2013,Niemela:JFM2001,Horn:JFM2013} for (a) $\mathrm{Pr}<1$, and (b) $\mathrm{Pr}\geq 1$. The error bars (shown only for our datasets) represent the standard deviation of the dataset with respect to the temporal average.}
	\label{fig:Model_Re}
\end{figure}
\begin{figure}[hbtp]
	\includegraphics[scale=0.39]{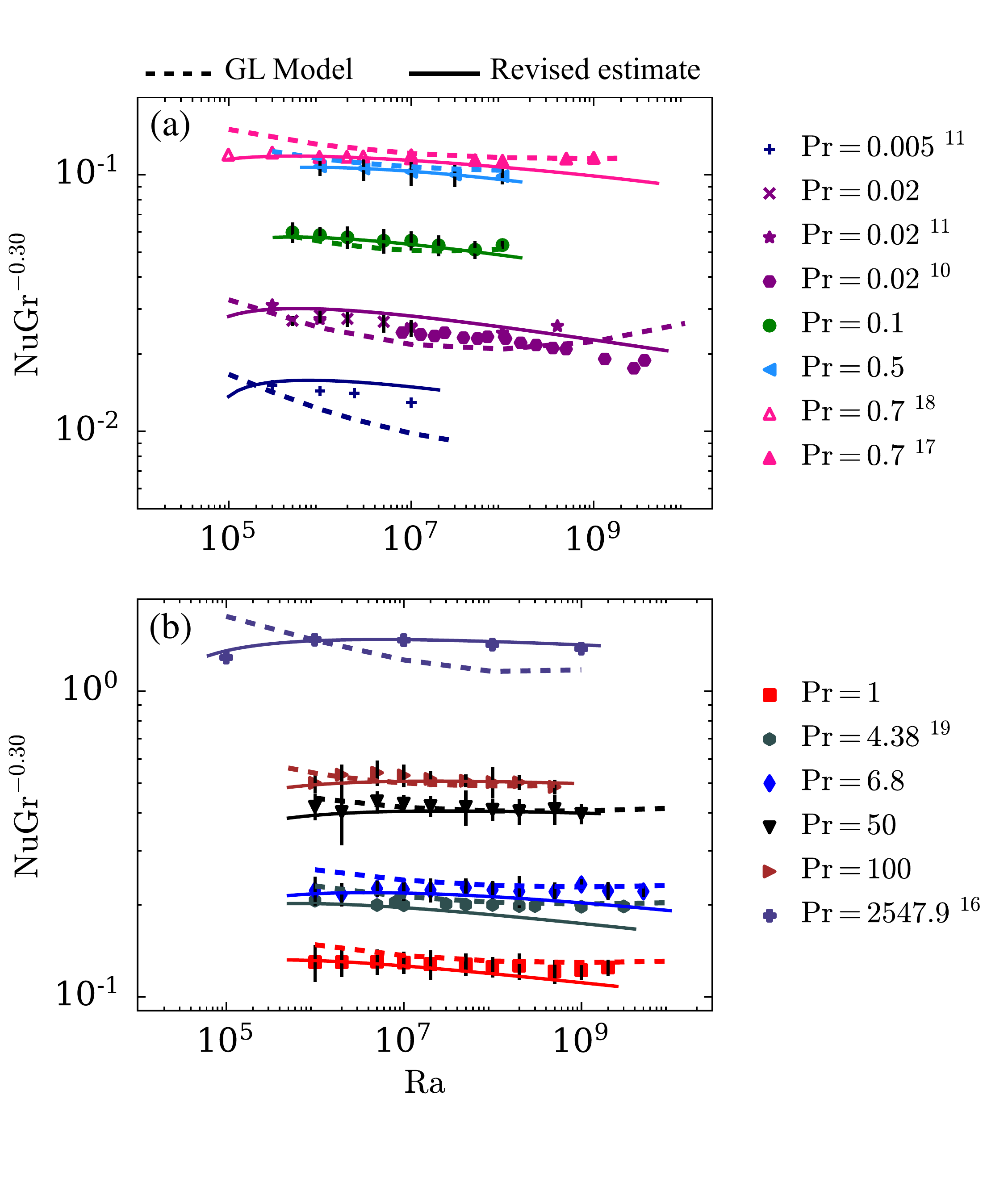}
	\caption{(color online) Comparison between the predictions of Nu vs. Ra using {\color{black}the original GL model (dashed curves) and our proposed modifications (solid curves)} with the results from our work and from the literature~\cite{Scheel:PRF2017,Cioni:JFM1997,Emran:JFM2008,Wagner:PF2013,Kaczorowski:JFM2013,Horn:JFM2013} for (a) $\mathrm{Pr}<1$, and (b) $\mathrm{Pr}\geq 1$. The error bars (shown only for our datasets) represent the standard deviation of the dataset with respect to the temporal average.}
	\label{fig:Model_Nu}
\end{figure}
\begin{figure}[hbtp]
	\includegraphics[scale=0.39]{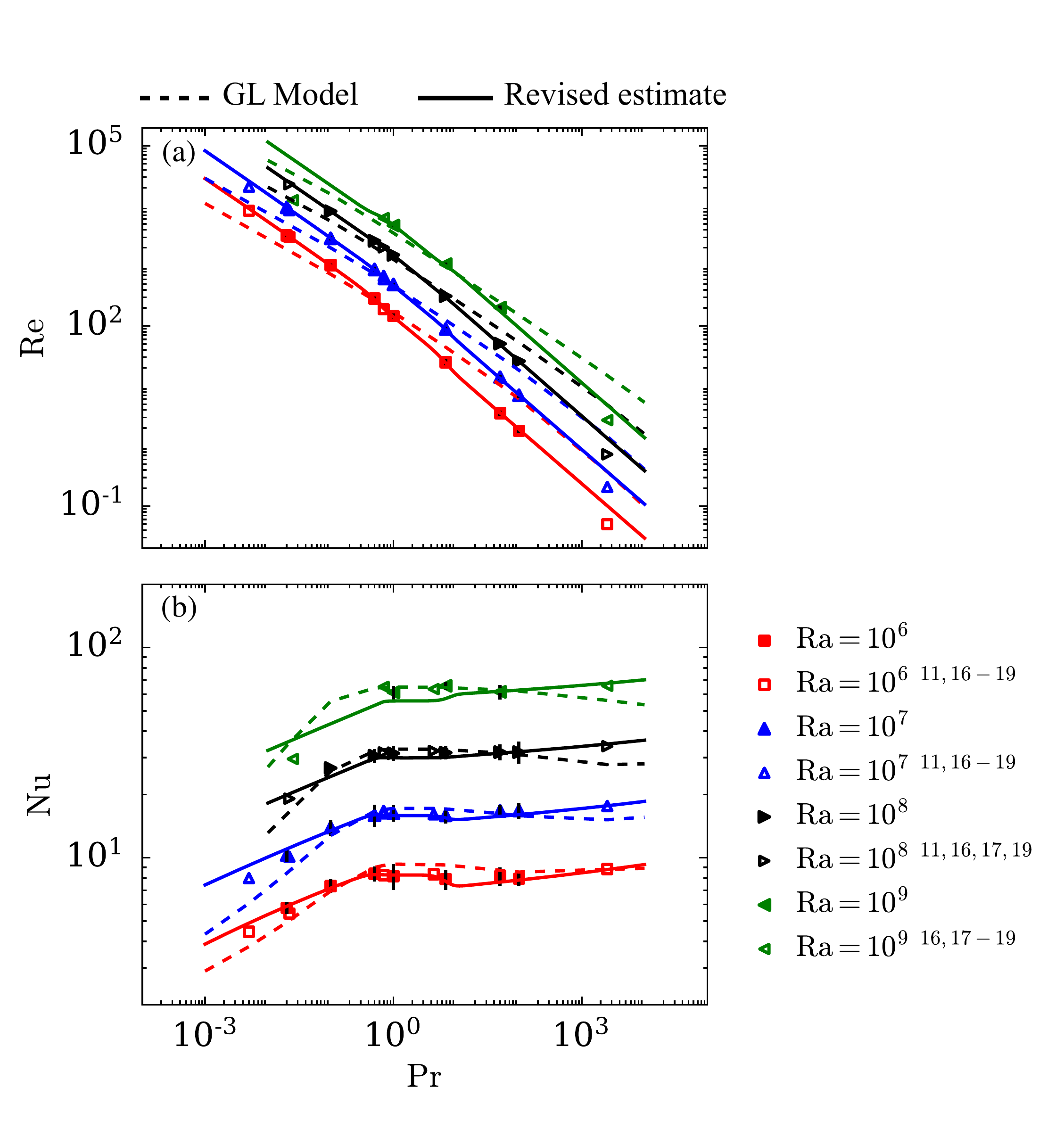}
	\caption{(color online) Comparison between the predictions of (a) Re and (b) Nu vs. Pr using {\color{black}the original GL model (dashed curves) and our proposed modifications (solid curves)} with the results from our work (filled markers) and from Refs.~\cite{Scheel:PRF2017,Cioni:JFM1997,Emran:JFM2008,Wagner:PF2013,Kaczorowski:JFM2013,Horn:JFM2013} (unfilled markers). The error bars (shown only for our datasets) represent the standard deviation of the dataset with respect to the temporal average.}
	\label{fig:Model_ReNu_Pr}
\end{figure}

In Fig.~\ref{fig:Model_Re}(a,b), we plot the normalized Reynolds number, $\mathrm{Re}\mathrm{Ra}^{-0.5}$, computed using our simulation data  and those of Refs.~\cite{Scheel:PRF2017,Cioni:JFM1997,Emran:JFM2008,Wagner:PF2013,Niemela:JFM2001,Horn:JFM2013}, versus Ra. To avoid clutter,  we exhibit the results for $\mathrm{Pr} <1$ in Fig.~\ref{fig:Model_Re}(a) and those for $\mathrm{Pr} \geq 1$ in Fig.~\ref{fig:Model_Re}(b). {\color{black}The dashed and the solid curves in Fig.~\ref{fig:Model_Re} denote Re predicted by the GL model and our revised estimates respectively.}
From the above figure and Table~\ref{table:Model_Comparison}, it is clear that the revised {\color{black} estimates of Re} are in better agreement with the observed results compared to the original GL model, especially for extreme Prandtl numbers. 
Further, the trend of {\color{black}estimated Re} is also in better agreement with the numerical and experimental results. (Note that the trend of Re computed based on different large-scale velocities does not change even though there may be minor differences in absolute values~\cite{Ahlers:RMP2009}). 
{\color{black}This improvement in the estimation of Re is crucial because} the predictions of Re are more sensitive to modeling parameters compared to Nu due to a larger range of the scaling exponent.

In Figs.~\ref{fig:Model_Nu}(a,b), we plot the normalized Nusselt number,  $\mathrm{NuGr^{-0.3}}$, computed using our simulation data  along with those of Refs.~\cite{Scheel:PRF2017,Cioni:JFM1997,Emran:JFM2008,Wagner:PF2013, Kaczorowski:JFM2013,Horn:JFM2013}, versus Ra.  We employ the Grashoff number   $\mathrm{Gr} = \mathrm{Ra}/\mathrm{Pr}$ in the  $ y $  axis to avoid clutter; this is because  $\mathrm{Nu} \sim \mathrm{Ra}^{0.3}$ (with a weak dependence on Pr).  
These figures, along with Table~\ref{table:Model_Comparison}, indicate that the {\color{black}revised estimates of Nu (solid curves) are more accurate compared to those predicted by the original GL model (dashed curves).}
It is interesting to note that for extreme Prandtl numbers ($\mathrm{Pr}=0.005$, $2547.9$), {\color{black}the accuracy of the revised estimates of Nu is significantly improved} with only $2.3\%$ deviation from the actual values for $\mathrm{Pr}=2547.9$ and 9.6\% deviation for $\mathrm{Pr}=0.005$. Contrast this with the GL model, where we observe 17\% deviation for both $\mathrm{Pr}=2547.9$ and $0.005$. For $\mathrm{Pr} \sim 1$, the {\color{black}accuracy of the revised estimates of Nu and those predicted by the GL model are comparable, with the former being more accurate for $\mathrm{Ra} < 10^8$ but marginally less for larger Ra}. Thus, we observe an overall improvement in the predictions of Nu, though it is not as significant as it was for Re.

In Figs.~\ref{fig:Model_ReNu_Pr}(a,b), we contrast the Pr dependence on {\color{black}our estimates of Re and Nu and those of the GL model}. Here, we plot the {\color{black}predictions of $\mathrm{Re}(\mathrm{Pr})$ and $\mathrm{Nu}(\mathrm{Pr})$} along with the actual values computed using our data and those of Refs.~\cite{Scheel:PRF2017,Emran:JFM2008,Wagner:PF2013,Niemela:JFM2003,Kaczorowski:JFM2013,Horn:JFM2013}. We choose four Rayleigh numbers for our comparisons: $10^6$, $10^7$, $10^8$, and $10^9$. As expected based on our earlier discussions, {\color{black}the revised estimates of $\mathrm{Re}(\mathrm{Pr})$ are more accurate than those of the GL model [See Fig.~\ref{fig:Model_ReNu_Pr}(a)]. We also observe improvements in the predictions of Nu, especially for $\mathrm{Pr}\ll 1$ and $\mathrm{Pr} \gg 1$ [see Fig.~\ref{fig:Model_ReNu_Pr}(b)].} This is again consistent with our earlier discussions.  

{\color{black}The improvements thus in the predictions of Re and Pr underscore the importance of considering the additional Ra and Pr dependence on the scaling of the dissipation rates and the viscous boundary layers in convection.}

\subsection{Limiting cases: Power-law expressions}
\label{subsec:Limiting_cases}

\begin{figure}[t]
	\includegraphics[scale=0.4]{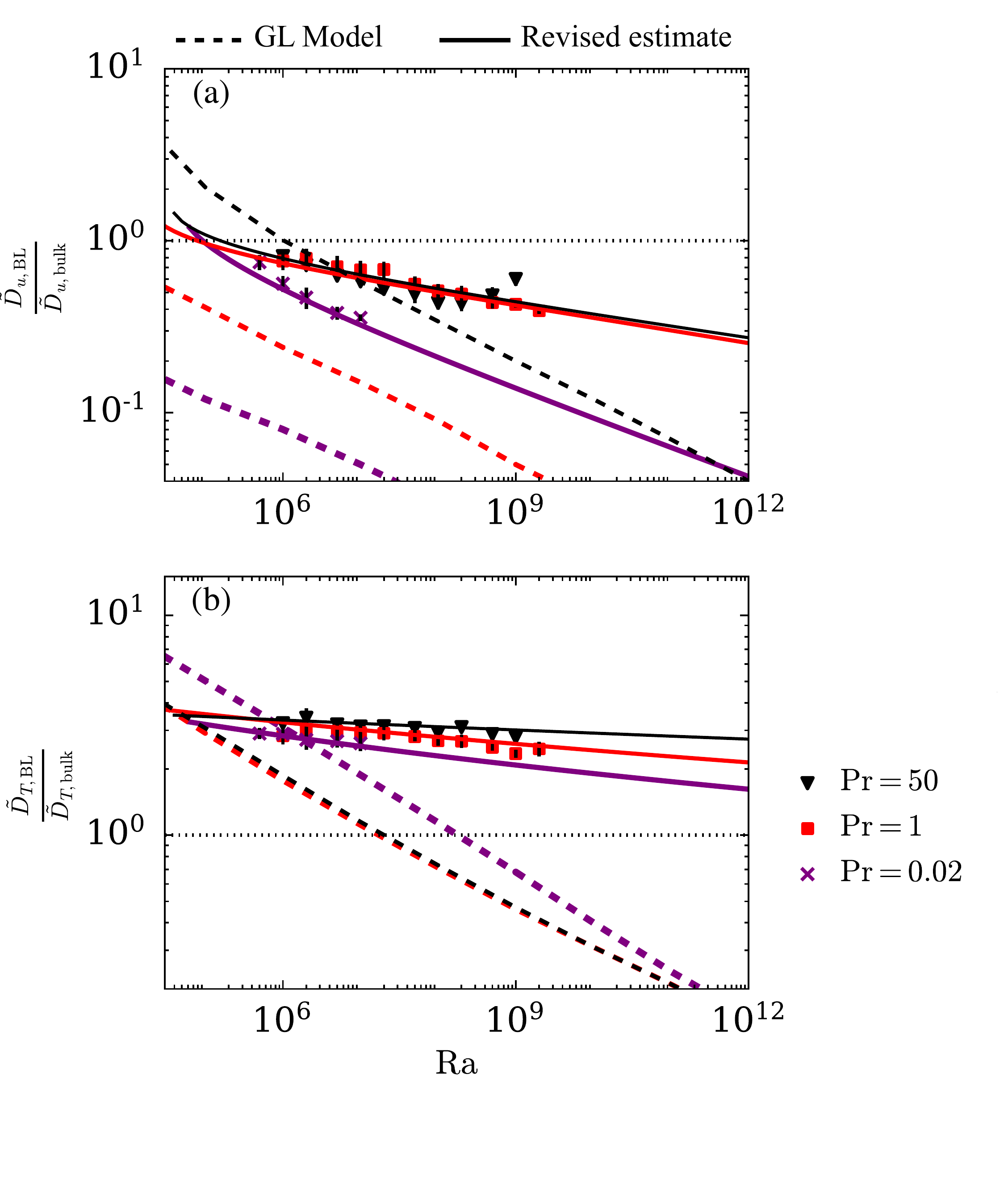}
	\caption{(color online) Estimates of (a) $\tilde{D}_{u,\mathrm{BL}}/\tilde{D}_{u,\mathrm{bulk}}$ and  (b) $\tilde{D}_{T,\mathrm{BL}}/\tilde{D}_{T,\mathrm{bulk}}$  using {\color{black}Eqs.~(\ref{eq:VD_ratio},  \ref{eq:TD_ratio})} (solid curves) and the GL model (dashed curves) for $\mathrm{Pr=0.02}$ (purple), $\mathrm{Pr}=1$ (red), and $\mathrm{Pr}=50$ (black). Points obtained from our simulation data are also displayed. The dotted horizontal lines in (a) and (b) represent $\tilde{D}_{u,\mathrm{BL}}/\tilde{D}_{u,\mathrm{bulk}}=1$ and $\tilde{D}_{T,\mathrm{BL}}/\tilde{D}_{T,\mathrm{bulk}}=1$ respectively. The error bars represent the standard deviation of the dataset with respect to the temporal average.}
	\label{fig:D_ratio}
\end{figure}

Recall from Sec.~\ref{sec:Governing_Equations} that {\color{black}Eqs.~(\ref{eq:Equation1}) and (\ref{eq:Equation2})} reduce to power-law scaling in the limiting cases:  $\tilde{D}_{u,\mathrm{bulk}} \gg \tilde{D}_{u,\mathrm{BL}}$ and $\tilde{D}_{u,\mathrm{bulk}} \ll \tilde{D}_{u,\mathrm{BL}}$. First, we will first estimate the regimes of Ra and Pr where the viscous and thermal dissipation rates dominate in the bulk or in the boundary layers.
Using $f_i$'s  and Eqs.~(\ref{eq:U^3/d}) to (\ref{eq:kappa Delta^2/d^2}), we deduce that
\begin{eqnarray}
\frac{\tilde{D}_{u,\mathrm{BL}}}{\tilde{D}_{u,\mathrm{bulk}}} &=& \frac{f_2}{f_1} \frac{d}{\delta_u}\frac{1}{\mathrm{Re}}, \label{eq:VD_ratio}\\
\frac{\tilde{D}_{T,\mathrm{BL}}}{\tilde{D}_{T,\mathrm{bulk}}} &=& \frac{2f_4}{f_3} \mathrm{\frac{Nu}{RePr}}. \label{eq:TD_ratio}
\end{eqnarray}
In Figs.~\ref{fig:D_ratio}(a,b), we exhibit the plots of the above estimates for $\mathrm{Pr=0.02}$, $1$, and $50$. We also exhibit the numerically computed points in the same figure; these points are consistent with the estimates given by Eqs.~(\ref{eq:VD_ratio}) and (\ref{eq:TD_ratio}). On the other hand, the ratio of the dissipation rates estimated using the GL model [by employing the bulk and the boundary layer terms of Eqs.~(\ref{eq:GL1}) and (\ref{eq:GL2})] deviate significantly from the numerically computed points. 

The plots show that the thermal dissipation rate in the boundary layers exceeds that in the bulk by a factor of two to four  for all Pr. On the other hand, the viscous dissipation rate in the bulk exceeds that in the boundary layers for $\mathrm{Ra} \gtrapprox 10^5$. These observations are in agreement with previous studies~\cite{Bhattacharya:PF2018,Bhattacharya:PF2019}.  The plots imply that $\tilde{D}_{u,\mathrm{BL}}$ dominates $\tilde{D}_{u,\mathrm{bulk}}$ only for $\mathrm{Ra} \ll 10^5$, where $\mathrm{Nu} \approx 1$. However, recall that the power-law relations for this limiting case, given by Eqs.(\ref{eq:Re_Viscous}) and (\ref{eq:Nu_Viscous}), are invalid for small Nu. Thus, we do not examine this limiting case further.
 
For the regimes characterized by $\tilde{D}_{u,\mathrm{bulk}} \gg \tilde{D}_{u,\mathrm{BL}}$, we plug the best-fit relation for $f_i$ in Eqs.~(\ref{eq:Re_Turbulent}) and (\ref{eq:Nu_Turbulent}) to obtain the following:
\begin{eqnarray}
\mathrm{Re} &=& 
\begin{cases}
0.76 \mathrm{Ra}^{0.42}\mathrm{Pr}^{-0.72}, \quad \mbox{Small Pr}, \\
0.20 \mathrm{Ra}^{0.50}\mathrm{Pr}^{-0.86}, \quad \mbox{Moderate Pr}, \\
0.11 \mathrm{Ra}^{0.55}\mathrm{Pr}^{-0.94}, \quad \mbox{Large Pr},
\end{cases}
\label{eq:Re_turbulent_actual} \\
\mathrm{Nu} &=&
\begin{cases}
0.30 \mathrm{Ra}^{0.27}\mathrm{Pr}^{0.11}, \quad \mbox{Small Pr}, \\
0.21 \mathrm{Ra}^{0.29}\mathrm{Pr}^{-0.03}, \quad \mbox{Moderate Pr}, \\
0.21 \mathrm{Ra}^{0.30}\mathrm{Pr}^{-0.03}, \quad \mbox{Large Pr}.
\end{cases}
\label{eq:Nu_turbulent_actual}
\end{eqnarray}
Since $f_4$ is a very weak function of Ra and Pr, we assume it to be a constant ($\approx 0.37$). The Ra dependence described by Eqs.~(\ref{eq:Re_turbulent_actual}) and (\ref{eq:Nu_turbulent_actual}) is consistent with the scaling observed for large Rayleigh numbers ($10^8 \ll \mathrm{Ra} \ll 10^{12}$) in the literature.~\cite{Scheel:PRF2017,Castaing:JFM1989,Qiu:PRE2002,Brown:JSM2007,Emran:JFM2008,Wagner:PF2013,Nikolaenko:JFM2005,Verzicco:JFM2003, Scheel:JFM2012,Scheel:JFM2014,Pandey:PF2016, Pandey:PRE2016,Horn:JFM2013} Further, the above relation for Re and Nu in the small Pr regime is not very far from GL's predictions of $\mathrm{Re} \sim \mathrm{Ra}^{2/5}\mathrm{Pr}^{-3/5}$ and $\mathrm{Nu} \sim  \mathrm{Ra}^{1/5}\mathrm{Pr}^{1/5}$.  
The derived relation for Nu is also in agreement with analytically derived upper bounds of
 $\mathrm{Nu} \lesssim \mathrm{Ra}^{1/3} \ln(\mathrm{Ra}^{2/3})$~\cite{Constantin:JSP1999} and $\mathrm{Nu} \leq 0.644 \mathrm{Ra}^{1/3} \ln(\mathrm{Ra}^{1/3})$.~\cite{Doering:JFM2006} Equation~(\ref{eq:Nu_turbulent_actual}) also  suggests that Nu is a weak function of Pr for moderate and large Pr [see Fig.~\ref{fig:Model_ReNu_Pr}(b)]. 

For very large Ra ($\gg 10^{12}$), some recent works~\cite{Iyer:PNAS2020,Zhu:PRL2018} reveal that the Nusselt number  scales in the band $\mathrm{Ra}^{0.33}$ to $\mathrm{Ra}^{0.35}$. Unfortunately, {\color{black}our predictions are not very accurate in this regime}; this is because {\color{black}the functional forms of $f_i$ are constructed using data from simulations} with $\mathrm{Ra} \lessapprox 10^{10}$. Note that for larger Ra, we expect the suppression of viscous and thermal dissipation rate to weaken because of the thin boundary layers. This can, in turn, cause the scaling exponent for Nu to increase. For example,  $f_1$ and $f_3$ may scale as
\begin{equation}
f_1 \sim \mathrm{Ra}^{-0.14}, \quad f_3 \sim \mathrm{Ra}^{-0.16},
\label{eq:fi_ultimate}
\end{equation}
 instead of $\mathrm{Ra}^{-0.21}$ as per Eqs.~(\ref{eq:c1_Fit}) and (\ref{eq:c3_Fit}). Plugging the above expressions for $f_1$ and $f_3$ in Eq.~(\ref{eq:Nu_Turbulent}) gives 
 \begin{equation}
 \mathrm{Nu} \sim \mathrm{Ra}^{0.33}, \nonumber
 \end{equation}
  which is consistent with the results of \citet{Iyer:PNAS2020}. However, the scalings for $f_1$ and $f_3$, given by Eq.~(\ref{eq:fi_ultimate}), are  conjectures that need to be verified using simulations with large Ra's.  In a future work, we plan to upgrade {\color{black}our present work} by taking inputs from large Ra simulations.

We conclude in the next section.
 \section{Conclusions} \label{sec:Conclusions}
  In this paper, we {\color{black}enhance} Grossmann and Lohse's model to provide improved predictions of Reynolds and Nusselt numbers in turbulent Rayleigh-Bénard convection. The process of obtaining this relation involves Grossman and Lohse’s idea of splitting the total viscous and thermal dissipation rates into bulk and boundary layer contributions and using the exact relations of Shraimann and Siggia. In the {\color{black}present work, we address the additional Ra and Pr dependence on} the viscous and thermal dissipation rates in the bulk compared to free turbulence, as well as the deviation of viscous boundary layer thickness from Prandtl-Blasius theory. 

{\color{black}The Reynolds and Nusselt numbers are obtained by solving a cubic polynomial equation consisting of} four functions $f_i(\mathrm{Ra,Pr})$ that are prefactors for the dissipation rates in the bulk and boundary layers. {\color{black}Note these prefactors were constants in the original GL model.
The aforementioned }functions  are determined using machine learning (regression analysis) on 60 datasets obtained from direct numerical simulations of RBC. The cubic polynomial equation reduces to power-law expressions in the limit of viscous dissipation rate dominating in the bulk.
 	
{\color{black} Using functional forms for the prefactors for the dissipation rates improves the predictions for both Re and Nu compared to the GL model. We observe significant improvements in the predictions of Re,} which is important because Re is more sensitive to modeling parameters compared to Nu. {\color{black}The improvement in the predictions of Nu} is more pronounced for extreme Pr regimes ($\mathrm{Pr} \leq 0.02$ and $\geq 100$). {\color{black}Our results underscore the importance of applying data-driven methods to improve existing models, a practice that has recently been picking up pace in research on turbulence~\cite{Pandey:JoT2020,Parish:JCP2016}. Presently, our work} takes inputs from data that are restricted to $\mathrm{Ra}<10^{10}$ and unit aspect ratio. {\color{black}Our predictions can be further enhanced} after determining $f_i$ for $\mathrm{Ra} > 10^{10}$ and for different aspect ratios. {\color{black}Moreover, our work can be extended to convection with magnetic fields following the approach of Z{\"u}rner \textit{et al}~\cite{Zuerner:PRE2016,Zuerner:PF2020}.}
 
 We believe that {\color{black}our results} will be valuable to the scientific and engineering community, especially where flows with extreme Prandtl numbers are involved. For example, {\color{black}they} will help understand the fluid dynamics and heat transport in liquid metal batteries which involve small Pr convection.~\cite{Kelley:AMR2018} On the other end, {\color{black}our analysis} will help strengthen our knowledge on mantle convection, which involves large Pr flow~\cite{Ahlers:RMP2009,Chilla:EPJE2012,Prakash:CES2017}. This will, in turn, enable us to make better predictions of seismic disturbances and the earth's magnetic field. Apart from this, our {\color{black}present work} should also aid in expanding our knowledge on oceanic and atmospheric flows and thus enable us to make improved weather predictions.

\section*{Acknowledgements}
The authors thank Arnab Bhattacharya, K. R. Sreenivasan, J{\"o}rg Schumacher, and Ambrish Pandey for useful discussions. The authors acknowledge Roshan Samuel, Ali Asad, Soumyadeep Chatterjee, and Syed Fahad Anwer for their contributions to the development of the finite-difference solver SARAS. Our numerical simulations were performed on Shaheen II of {\sc Kaust} supercomputing laboratory, Saudi Arabia (under the project k1416) and on HPC2013 of IIT Kanpur, India. 

\section*{Data availability}
The data that support the findings of this study are available from the corresponding author upon reasonable request.

\appendix
{\color{black}
\section{Robustness of the estimated parameter values for $f_i(\mathrm{Ra,Pr})$}
\label{sec:Robustness}
In this section, we check the robustness of the parameter values for $f_i(\mathrm{Ra,Pr})$ estimated in Sec.~\ref{subsec:ML}. Towards this objective, we employ regression algorithm on a reduced training set consisting of 30 data-points, which is half of the total number of data-points, and test the algorithm on the remaining 30 data-points. Starting from the point corresponding to $\mathrm{Pr}=0.02$, $\mathrm{Ra}=5 \times 10^5$, we take alternate data-points from Tables~\ref{table:SimDetails} and \ref{table:SimDetails_Pr>1} for training and the remaining data-points for testing. We obtain the following fits for $f_i$ for the reduced training set:
\begin{eqnarray}
f_1 &=& 0.72 H_1\mathrm{Pr}^{0.30} + 28 H_2 \mathrm{Ra}^{-0.21}\mathrm{Pr}^{0.52} \nonumber \\ 
&& + 150 H_3 \mathrm{Ra}^{-0.33}\mathrm{Pr}^{0.79}, \label{eq:c1_Fit_Red} \\
\frac{f_2}{\delta_u} &=& 4.1 H_1 \mathrm{Ra}^{0.26}\mathrm{Pr}^{-0.27} + 6.9 H_2 \mathrm{Ra}^{0.23}\mathrm{Pr}^{-0.30}  \nonumber \\
&&+ 21 H_3 \mathrm{Ra}^{0.15}\mathrm{Pr}^{-0.18}, \label{eq:c2p_Fit_Red} \\
f_3 &=& 0.087 H_1\mathrm{Ra}^{-0.14}\mathrm{Pr}^{-0.16} 
{\color{black}+} 0.26 H_2\mathrm{Ra}^{-0.21}\mathrm{Pr}^{-0.17} \nonumber \\
&&+ 0.40 H_3\mathrm{Ra}^{-0.24}\mathrm{Pr}^{-0.095}, \label{eq:c3_Fit_Red} \\
f_4 &=& 0.45 H_1\mathrm{Ra}^{-0.012}\mathrm{Pr}^{0.0075} 
+ 0.42 H_2\mathrm{Ra}^{-0.0078}\mathrm{Pr}^{0.0050} \nonumber \\
&&+ 0.36 H_3\mathrm{Pr}^{0.0161}. \label{eq:c4_Fit_Red}
\end{eqnarray}
We observe that the fits given by Eqs.~(\ref{eq:c1_Fit_Red}-\ref{eq:c4_Fit_Red}) are similar to Eqs.~(\ref{eq:c1_Fit}-\ref{eq:c4_Fit}), which correspond to the fits obtained when all the datapoints were used as training sets. The average deviation between the $f_i$'s predicted by the fits and the actual values of the test set are 25\%, 20\%, 13\%, and 71\% for $f_1$, $f_2/\delta_u$, $f_3$, and $f_4$ respectively. These deviations are almost the same as those observed when all the datasets were used for training and testing. Further, if we train our algorithm using only 15 datasets (every fourth set from Tables~\ref{table:SimDetails} and \ref{table:SimDetails_Pr>1}), we obtain
\begin{eqnarray}
f_1 &=& 0.68 H_1\mathrm{Pr}^{0.31} + 25 H_2 \mathrm{Ra}^{-0.20}\mathrm{Pr}^{0.47} \nonumber \\ 
&& + 238 H_3 \mathrm{Ra}^{-0.37}\mathrm{Pr}^{0.81}, \label{eq:c1_Fit_RedRed} \\
\frac{f_2}{\delta_u} &=& 3.7 H_1 \mathrm{Ra}^{0.26}\mathrm{Pr}^{-0.27} + 5.8 H_2 \mathrm{Ra}^{0.24}\mathrm{Pr}^{-0.33}  \nonumber \\
&&+ 23 H_3 \mathrm{Ra}^{0.15}\mathrm{Pr}^{-0.19}, \label{eq:c2p_Fit_RedRed} \\
f_3 &=& 0.060 H_1\mathrm{Ra}^{-0.12}\mathrm{Pr}^{-0.17} 
{\color{black}+} 0.23 H_2\mathrm{Ra}^{-0.20}\mathrm{Pr}^{-0.19} \nonumber \\
&&+ 0.40 H_3\mathrm{Ra}^{-0.24}\mathrm{Pr}^{-0.090}, \label{eq:c3_Fit_RedRed} \\
f_4 &=& 0.42 H_1\mathrm{Ra}^{-0.0099} 
+ 0.41 H_2\mathrm{Ra}^{-0.0069}\mathrm{Pr}^{0.0059} \nonumber \\
&&+ 0.38 H_3, \label{eq:c4_Fit_RedRed}
\end{eqnarray}
with the average deviation between the $f_i$'s predicted by the fits and the actual values of the test set being 26\%, 21\%, 16\%, and 71\% for $f_1$, $f_2/\delta_u$, $f_3$, and $f_4$ respectively. We observe that there are visible changes in the parameter values estimated using 15 datasets.
Thus, we infer that the parameter values estimated using more than 30 datasets are reasonably robust.
}
\section*{References}
%

\end{document}